\documentclass[12pt]{article}
\pdfoutput=1
\usepackage[english]{babel}
\usepackage[latin1]{inputenc}

\usepackage[intlimits]{amsmath}
\usepackage{amssymb}

\usepackage{url}
\usepackage{graphicx}
\usepackage{color}

\numberwithin{equation}{section}

\def\lsim{\mathrel{\raise.3ex\hbox{$<$\kern-.75em\lower1ex\hbox{$\sim$}}}}
\def\gsim{\mathrel{\raise.3ex\hbox{$>$\kern-.75em\lower1ex\hbox{$\sim$}}}}

\newcommand{\MeV}{\,\text{MeV}}
\newcommand{\GeV}{\,\text{GeV}}
\newcommand{\TeV}{\,\text{TeV}}
\newcommand{\cm}{\,\text{cm}}

\newcommand{\s}{\,\text{s}}

\newcommand{\kpc}{\,\text{kpc}}

\newcommand{\be}{\begin{equation}}
\newcommand{\ee}{\end{equation}}
\newcommand{\bea}{\begin{eqnarray}}
\newcommand{\eea}{\end{eqnarray}}

\newcommand{\picwid}{0.9}

\begin{document}

\date{\mbox{ }}

\title{
\vspace{-3.1cm}
{\begin{flushright}
{\normalsize
ULB-TH/09-44\\
TUM-HEP 744/09\\
DESY 09-221\\
} 
\end{flushright}}
\vspace{1.2cm}
\bf Intense Gamma-Ray Lines from Hidden Vector Dark Matter Decay \\[6mm]}

\author{Chiara Arina$^a$, Thomas Hambye$^a$,
Alejandro Ibarra$^b$, Christoph Weniger$^{c}$\\[3mm]
{\normalsize\it a Service de Physique Th\'eorique,}\\[-0.15cm]
{\it\normalsize Universit\'e Libre de Bruxelles, 1050 Brussels, Belgium} \\[1.5mm]
{\normalsize\it b  Physik-Department T30d, Technische Universit\"at M\"unchen,}\\[-0.15cm]
{\it\normalsize James-Franck-Stra\ss{}e, 85748 Garching, Germany}\\[1.5mm]
{\normalsize\it c Deutsches Elektronen-Synchrotron DESY, Hamburg}\\[-0.15cm]
{\it\normalsize Notkestra\ss{}e 85, 22603 Hamburg, Germany}\\[1.5mm]
}
\maketitle

\thispagestyle{empty}

\begin{abstract}
\noindent
Scenarios with hidden, spontaneously broken, non-abelian gauge groups 
contain a natural dark matter candidate, the hidden vector, whose longevity 
is due to an accidental custodial symmetry in the renormalizable
Lagrangian. Nevertheless,  non-renormalizable dimension six operators
break the custodial symmetry and induce the decay
of the dark matter particle at cosmological times. We discuss in this paper the cosmic
ray signatures of this scenario and we show
that the decay of hidden vector dark matter particles generically
produce an intense gamma ray line which could be observed
by the Fermi-LAT experiment, if the scale of custodial symmetry breaking is close
to the Grand Unification scale. This gamma line proceeds directly from a tree level
dark matter 2-body decay in association with a Higgs boson.
Within this model we also perform a determination of the relic density
constraints taking into account the dark matter
annihilation processes with one dark matter particle in the final state. The corresponding direct detection rates can be easily of order the current experimental sensitivities.
\end{abstract}

\newpage

\section{Introduction}
\label{Intro}

One of the most striking features of the dark matter (DM) particle is its longevity
at cosmological scales. This fact could be accommodated {\it ad hoc} by
imposing a new symmetry (discrete or continuous) which prevents the decay of
the dark matter particle, such as the $R$-parity in the Minimal Supersymmetric
Standard Model or the $Z_2$ symmetry assumed in many 
phenomenological models. The longevity
of the dark matter particle is more elegantly explained, however, if it arises
as the result of an accidental symmetry of the Lagrangian, in complete analogy
to the longevity of the proton, which in the Standard Model framework is
explained by being the proton the lightest particle carrying baryon number. A
simple implementation of this idea consists in extending the Standard Model
gauge group with a non-abelian gauge symmetry, under which all the Standard
model particles are singlets, which is spontaneously broken by the vacuum
expectation value of a standard model singlet scalar particle.
The renormalizable part of the Lagrangian of
this model presents an accidental custodial symmetry which prevents the decay
of the hidden vector bosons, thus predicting the existence of a particle with
the correct dark matter properties~\cite{Hambye:2008bq,Hambye:2009fg}. 

The simplest example of such class of models introduces in the hidden sector
an extra $SU(2)_{\rm HS}$ gauge group plus a complex scalar doublet of this
gauge symmetry, which acquires a vacuum expectation value.  After the
$SU(2)_{\rm HS}$ spontaneous symmetry breaking, the renormalizable part of the
Lagrangian presents a $SO(3)$ custodial symmetry which makes the three
components of the $SU(2)_{\rm HS}$ vector boson degenerate in mass and stable.
For wide ranges of the parameters of the model, the relic
abundance of the vector multiplets can reproduce the observed dark matter
abundance. Furthermore, these parameters are also consistent with the present
constraints from electroweak precision measurements and from direct dark
matter searches.

Being the $SO(3)$ custodial symmetry an accidental symmetry it is plausibly
broken explicitly by higher dimensional operators in the Lagrangian. This is
again in complete analogy with the baryon number violating dimension six
operators that necessarily appear in the Standard Model Lagrangian, unless the
baryon number conservation arises as a residual symmetry of an underlying gauge group.
Indeed, there are dimension six operators which violate the custodial symmetry
which can induce the decay of the dark matter particle, whereas analogous
dimension five operators are absent.\footnote{In this respect, to consider an accidental symmetry
is different from justifying the dark matter stability from a residual discrete subgroup of 
a Grand Unification gauge group \cite{Mohapatra:1986su,Martin:1992mq,Aulakh:1997fq,Aulakh:2000sn,Kadastik:2009dj,Frigerio:2009wf}, where the latter forbids any decay.}  The scale of custodial symmetry breaking
has a lower bound stemming from the requirement that the dark matter lifetime
has to be longer than the age of the Universe, $\tau_{\rm DM}\gsim 10^{17}\s$.
Furthermore, the dark matter decay produces a flux of stable particles, such
as positrons, antiprotons and gamma rays.  The requirement that the exotic
flux of cosmic rays does not exceed the measured fluxes translates into a more
stringent constraint on the dark matter lifetime
and in turn on the scale of custodial symmetry breaking.

On the other hand, a series of experiments measuring high-energy charged cosmic rays
have recently reported strong indications for the existence of an excess of
positrons at high energies. Namely, the PAMELA measurements of the positron
fraction show an energy spectrum which rises steeply at energies 7-100 GeV,
possibly extending towards higher energies~\cite{A08}, while the secondary 
positron flux calculated from state-of-the-art propagation 
models~\cite{Moskalenko:1997gh}, together with the total electron plus 
positron flux measured by the Fermi collaboration~\cite{Abdo:2009zk}, predict a
positron fraction which decreases monotonically with the energy. Furthermore,
the electron plus positron flux measured by Fermi is harder than expected from
conventional diffusive models~\cite{Grasso:2009ma}, also suggesting the 
existence of an excess of electrons and positrons at higher energies, 
with a cut-off at around 1 TeV as observed by the 
H.E.S.S.~collaboration~\cite{A08a,Aharonian:2009ah}. It is interesting
to mention that one possible explanation for the electron/positron excesses 
is precisely the decay of dark matter particles in the Milky Way halo 
with a mass in the TeV range and a lifetime 
$\sim 10^{26}\s$~\cite{Ibarra:2008jk,Ibarra:2009dr}. 
Nevertheless, irrespectively
of the origin of the electron/positron excesses, these measurements
set a constraint on the exotic flux of electrons/positrons.
Furthermore, models of dark matter decay are severely constrained 
by the PAMELA measurements of the antiproton-to-proton ratio~\cite{A09},
which does not show a deviation from the expectations of conventional
production mechanism. 
Interestingly, for TeV mass dark matter particles which decay via
dimension six operators, this value for the lifetime naturally arises
if the dimension six  operators are suppressed by a mass scale close to the 
scale of grand 
unification~\cite{Eichler:1989br,Nardi:2008ix,Arvanitaki:2008hq,Hamaguchi:2008ta,Ruderman:2009ta,Ruderman:2009tj}. Therefore, present cosmic ray observations provide a way of testing some scenarios of Grand Unification.

One of the well-known smoking gun signals for dark matter in the sky is the
possible observation of a sharp gamma-ray line~\cite{Bergstrom:1988fp,Rudaz:1989ij,Bouquet:1989sr}.
The $\gamma\gamma$ and $\gamma Z$ lines have been exhaustively studied in
supersymmetric models, for the neutralino dark matter~\cite{Boudjema:2005hb,
Bergstrom:1997fh, Bern:1997ng, Ullio:1997ke, Bergstrom:1997fj,
Bergstrom:2005ss}, in the inert doublet model~\cite{Gustafsson:2007pc} and for
Kaluza-Klein dark matter~\cite{Bergstrom:2004nr}. Recently it has been pointed
out that a monochromatic gamma line can be also produced accompanied to an
Higgs boson~\cite{Jackson:2009kg}. Note that the gamma lines in all these
models are induced by annihilation of dark matter particles at one loop level
(for an exception see Ref.~\cite{Dudas:2009uq}).
However, in many situations the disentanglement of the gamma lines from the
background requires either a dark matter mass in the TeV range or
astrophysical boost factors to make the signal strong enough.  On the other
hand, intense gamma-ray lines can also appear at tree level in the decay of
dark matter gravitinos in supersymmetric scenarios where R-parity conservation
is not
imposed~\cite{Buchmuller:2007ui,Ibarra:2007wg,Ishiwata:2008cu,Buchmuller:2009xv,Choi:2009ng}.
In both cases, the observation of a gamma-ray line would be a very clean
indirect detection signature for annihilating or decaying dark matter and is a
promising signature to search for.

In this paper we will analyze the cosmic ray signatures stemming from the
decay of a hidden $SU(2)$ vector boson. A decay, unlike most annihilations processes, 
does not require any ``boost factor" in order to lead to cosmic ray rates well above backgrounds. 
Moreover it leads to larger extragalactic fluxes. We will work out the possible decay
modes in detail and concentrate on gamma-ray line and anti-matter signatures.
Most interestingly, we find that hidden vector dark matter decay
modes with gamma-ray lines in the final state are automatically present
already at tree level. These decay modes exist for every possible dimension
six operator leading to the hidden vector dark matter decay
and are hence a robust prediction of the model. 
Taking advantage of the fact that the dark matter has spin-1, the gamma lines arise directly at tree level from DM decays to $\gamma h$ or $\gamma \eta$ where $h$ and $\eta$ are the standard model and hidden sector Higgs boson respectively.
Furthermore, one
of the operators generates a kinetic mixing between hidden sector and the
hypercharge $U(1)_Y$ giving rise to two-body decay modes into charged leptons.
We will discuss these properties 
in light of recent and future cosmic-ray
observations.

The paper is organized as follows: in section 2 we will introduce the hidden
vector dark matter model and discuss the possible decay modes in detail. In
section 3 we will discuss these decay modes for the different operators
separately and show results for several benchmark models, giving particular
emphasis to gamma-ray line signatures. In section 4 we improve the calculation of the relic density taking into account the dark matter annihilations with one DM particle in the final state, and compute the direct detection cross sections it gives. We draw
conclusions in section 5.

\section{Hidden $SU(2)$ model with custodial symmetry breaking}

We consider an extension of the Standard Model where the gauge group contains
a hidden non-abelian group, $SU(2)_{\rm HS}$, with gauge bosons $A^\mu$. We
assume that this symmetry is spontaneously broken via the vacuum expectation
value of a complex $SU(2)_{\rm HS}$ doublet scalar field, $\phi$. We further assume that all the
Standard Model particles are singlets under $SU(2)_{\rm HS}$, thus the
Standard Model only couples to the hidden sector via the Higgs portal term
$|\phi|^2 |H|^2$, being $H$ the Standard Model Higgs doublet (note that the
kinetic mixing of the $SU(2)_{\rm HS}$  gauge multiplet $A^\mu$ with the SM
gauge bosons is forbidden by the non-abelian character of the extra gauge
symmetry).  Under these assumptions, the renormalizable part of the Lagrangian
reads:
\be
{\cal L}= {\cal L}^{SM} -\frac{1}{4} F^{\mu\nu} \cdot F_{\mu \nu}
+(\mathcal{D}_\mu \phi)^\dagger (\mathcal{D}^\mu \phi) -\lambda_m \phi^\dagger
\phi H^\dagger H-\mu^2_\phi \phi^\dagger \phi -\lambda_\phi (\phi^\dagger
\phi)^2 \,,
\label{inputlagr}
\ee
where $\mathcal{D}^\mu =\partial^\mu \phi - i\frac{g_\phi}{2} \tau \cdot
A^\mu$, being $\tau^a$, $a=1,2,3$ the generators of the hidden $SU(2)$ gauge group. If $\mu^2_\phi<0$, the hidden sector scalar field $\phi$ acquires a
vacuum expectation value, $v_\phi$, and the $SU(2)_{\rm HS}$ symmetry is
broken spontaneously,
with $v_\phi=(-\mu_{\phi}^2 \lambda+\lambda_m \mu^2/2)/(\lambda \lambda_\phi-\lambda_m^2/4)$. In the unitary $SU(2)_{\rm HS}$ gauge the Lagrangian of the theory is:
\begin{eqnarray}
  {\cal L}&=&{\cal L}_{SM}-\frac{1}{4} F_{\mu \nu} \cdot F^{\mu \nu}
  +\frac{1}{8} (g_\phi v_\phi)^2 A_\mu \cdot A^\mu+\frac{1}{8} g_\phi^2 A_\mu
  \cdot A^\mu \eta'^2 +\frac{1}{4} g_\phi^2 v_\phi A_\mu \cdot A^\mu \eta'
  \nonumber\\ &&+\frac{1}{2}(\partial_\mu \eta')^2
  -\frac{\lambda_m}{2}(\eta'+v_\phi)^2 H^\dagger H- \frac{\mu_\phi^2}{2}
  (\eta'+v_\phi)^2 - \frac{\lambda_\phi}{4} (\eta'+v_\phi)^4 \,,
  \label{lagrzerov}
\end{eqnarray}
which gives $M_A=g_\phi v_\phi/2$ and where $\eta'$ is the hidden sector Higgs boson.
This Lagrangian has only 4 independent parameters, which can be taken as $g_{\phi}$, $v_{\phi}$, $\lambda_{\phi}$ and $\lambda_m$.

Once the electroweak sector is broken, the hidden sector $\eta'$ mixes with the standard model Higgs boson $h'$ through the Higgs portal interaction $\lambda_m$
\begin{equation}
\begin{array}{cc}
h'  = & \cos\beta \ h + \sin\beta \ \eta  \,,\\
\eta'  = & -\sin\beta \  h + \cos\beta \  \eta\,.
\end{array}
\label{physscal}
\end{equation}
The complete Lagrangian in the $h,\,\eta$ physical state basis can be found in Ref.~\cite{Hambye:2008bq}
as a function of $g_{\phi}$, $v_{\phi}$, $\lambda_\phi$ and $\lambda_m$, 
together with the corresponding expression for the mixing angle $\beta$.

The Lagrangian in Eq.~(\ref{lagrzerov}) has a remarkable property: it displays a $SO(3)$ custodial
symmetry in the $A_i^\mu$ component space, which prevents any decay to $SO(3)$
singlets (such as Standard Model particles or $\eta'$).  Consequently, if the
model is described just by the renormalizable Lagrangian, the
three $A^\mu_i$ components are degenerate in mass and are absolutely stable.
Nevertheless, since this $SO(3)$ global symmetry is accidental, one 
expects in the Lagrangian the existence of non-renormalizable operators
suppressed by a large scale $\Lambda$ which break the custodial symmetry. 
The complete list of operators with dimension smaller or equal than
six which lead, after the spontaneous symmetry breaking of $SU(2)_{\rm HS}$
and $SU(2)_L\times U(1)_Y$, to the breaking of the $SO(3)$ custodial symmetry
reads:
\bea
\label{eqn:opA}
&{\rm (A)}&~~~\frac{1}{\Lambda^2}\  \mathcal{D}_{\mu}\phi^{\dagger}\phi\ \mathcal{D}_{\mu}H^{\dagger}H \,,\\
\label{eqn:opB}
&{\rm (B)}&~~~\frac{1}{\Lambda^2}\  \mathcal{D}_{\mu}\phi^{\dagger}\phi\  H^{\dagger}\mathcal{D}_{\mu}H \,,\\
\label{eqn:opC}
&{\rm (C)}&~~~\frac{1}{\Lambda^2}\  \mathcal{D}_{\mu}\phi^{\dagger}\mathcal{D}_{\nu}\phi\ F^{\mu\nu Y} \,,\\
\label{eqn:opD}
&{\rm (D)}&~~~\frac{1}{\Lambda^2}\ \phi^{\dagger} F^a_{\mu\nu}\frac{\tau^a}{2}\phi F^{\mu\nu Y}\,.
\eea
In turn, the breaking of the custodial symmetry leads to the decay of the dark
matter hidden gauge bosons. Let us discuss for each case the dominant decay
modes:

\paragraph{Case A.}
After the spontaneous breaking of the gauge symmetries the non-renormalizable
part of the Lagrangian has two parts: ${\cal L}^{\rm NR}_A={\cal L}^{\rm
NR}_{A1}+{\cal L}^{\rm NR}_{A2}$. The first one reads:
\begin{equation}
  {\cal L}_{\rm A1}^{\rm NR}=\frac{1}{\Lambda^2}\Big(\frac{-i g_{\phi}}{4}
  A^3_{\mu} \big( \eta'\eta' + 2 \eta' v_{\phi} + v^2_{\phi} \big) \frac{1}{2}
  \big(h'\partial_{\mu}h'+v\partial_{\mu}h' \big)  \Big) +{\rm h.c.}\,,
  \label{lnr1A}
\end{equation}
which induces the decay of the dark matter particle into scalars, by means of Eqs.~(\ref{physscal}):
$A\rightarrow \eta\eta$, $A\rightarrow h\eta$ and $A\rightarrow hh$. The
corresponding decay rates are:
\begin{eqnarray}
  \Gamma(A\rightarrow \eta\eta) & = & \frac{1}{3}\frac{1}{16\pi}
  \frac{g^2_{\phi}}{256 \Lambda^4}\Big(\sin^2\beta v_{\phi}^2+ v_{\phi}v \sin
  2\beta\Big)^2 \frac{\sqrt{(M_A^2-4 M^2_{\eta})^3}}{M_A^2}\,,\nonumber\\
  \Gamma(A\rightarrow h\eta) & = & \frac{1}{3}\frac{1}{64\pi}
  \frac{g^2_{\phi}}{256 \Lambda^4}\Big(v^2_{\phi}\sin 2\beta+4 v v_{\phi}
  \cos^2\beta\Big)^2
  \frac{\sqrt{\lambda(M_A,M_h,M_{\eta})^3}}{M_A^5}\,,\nonumber\\
  \Gamma(A\rightarrow hh) & = & \frac{1}{3}\frac{1}{16\pi}
  \frac{g^2_{\phi}}{256 \Lambda^4}\Big(\cos^2\beta v_{\phi}^2 - v_{\phi}v \sin
  2\beta\Big)^2 \frac{\sqrt{(M_A^2-4 M^2_{h})^3}}{M_A^2}\,.
\end{eqnarray}
with $\lambda(M_A,m_1,m_2)=M_A^4+m_2^4+m_1^4-2(m_1^2+m_2^2)M_A^2-2m_1^2m_2^2$.

In addition, there is a second term in the non-renormalizable Lagrangian:
\begin{equation}
  {\cal L}_{\rm A2}^{\rm NR}=\frac{1}{\Lambda^2}\Big(\frac{-i g_{\phi}}{4}
  A^3_{\mu} \big( \eta'\eta' + 2 \eta' v_{\phi} + v^2_{\phi} \big) \frac{i e}{4}
  B_{\mu}\big(h'h' + 2vh'+v^2 \big)\Big)+{\rm h.c.}\,,
  \label{lnr2A}
\end{equation}
which induces decays into a gauge boson and the hidden sector and standard
model Higgs bosons, $A\rightarrow \gamma \eta$, $A\rightarrow Z \eta$,
$A\rightarrow \gamma h$ and $A\rightarrow Z h$, with rates:
\begin{eqnarray}
  \Gamma(A\rightarrow \gamma\eta) & = & \frac{1}{3}\frac{1}{16\pi} \frac{3 e^2
  g_{\phi}^2 \cos^2\theta_W (\cos\beta v_{\phi}v^2+\sin\beta v
  v^2_{\phi})^2}{64 \Lambda^4} \frac{(M_A^2-M_{\eta}^2)}{M_A^3}\,,\nonumber\\
  \Gamma(A\rightarrow Z\eta) & = & \frac{1}{3}\frac{1}{64\pi} \frac{ e^2
  g_{\phi}^2 \sin^2\theta_W (\cos\beta v_{\phi}v^2+\sin\beta v
  v^2_{\phi})^2}{64 \Lambda^4} \times\nonumber\\ & &
  \times\Big(8+\frac{(M_A^2-M_{\eta}^2+M_Z^2)^2}{M_A^2
  M_Z^2}\Big)\frac{\sqrt{\lambda(M_A,M_Z,M_{\eta})}}{M_A^3}\;.
\end{eqnarray}
The decay rates into $\gamma h$ and into $Z h$ are obtained using Eqs.~(\ref{physscal}) and $M_\eta\rightarrow M_h$.

\paragraph{Case B.}
At low energies the non-renormalizable Lagrangian for case B is ${\cal L}^{\rm
NR}_B={\cal L}^{\rm NR}_{A1}-{\cal L}^{\rm NR}_{A2}$, where ${\cal L}^{\rm
NR}_{A1}$ and ${\cal L}^{\rm NR}_{A2}$ are given in Eqs.~(\ref{lnr1A}) and~(\ref{lnr2A}) (note the change
of sign in the operator involving two gauge bosons compared to the case A).
Thus, the decay modes and rates are identical to the case A.

\paragraph{Case C.}
After the symmetry
breaking the Lagrangian for this dimension six operator reads as
\begin{equation}
  {\cal L}^{\rm NR}_C= \frac{1}{\Lambda^2}2 A^3_{\nu}
  \big(\partial_{\mu}B_{\nu}-\partial_{\nu}B_{\mu} \big) \partial_{\mu}\eta'
  (\eta'+v_{\phi})+{\rm h.c.}\,.
\end{equation} 
Remarkably it leads only to the following two body decays
\begin{eqnarray}
  \Gamma(A\rightarrow \gamma\eta) & = &
  \frac{1}{3}\frac{1}{16\pi}\frac{g^2_{\phi}v^2_{\phi} \cos^2\beta
  \cos^2\theta_W}{8 \Lambda^4
  M_A^3}\Big((M_A^2-M^2_{\eta})^3+\frac{M^2_{\eta}}{M_A^2}(M_A^4+M^4_{\eta})(M^2_A-M_{\eta}^2)\Big)\,,\nonumber\\
  \Gamma(A\rightarrow Z\eta) & = & \frac{1}{3}\frac{1}{16\pi}\frac{g_{\phi}^2
  v_{\phi}^2 \cos^2\beta \sin^2\theta_W}{16 \Lambda^4
  M_A^3}\sqrt{\lambda(M_A,M_Z,M_{\eta})}\times\nonumber\\
  &&\times\Big(2(M_A^2-M_{\eta}^2)^2-3M_{Z}^2(M_A^2-2
  M_{\eta}^2)-\frac{M_Z^4(M_Z-M_{\eta})^2}{M_A^2}\Big)\,.
\end{eqnarray}
together with the corresponding decay channels into $\gamma h$ and $Z h$, which can be derived
using Eqs.~(\ref{physscal}) and substituting $M_{\eta} \rightarrow M_h$.

\paragraph{Case D.}
The non-renormalizable part of the Lagrangian contains two terms: ${\cal
L}^{\rm NR}_D={\cal L}^{\rm NR}_{D_1}+{\cal L}^{\rm NR}_{D_2}$. It contains a
kinetic mixing term:
\begin{equation}
  {\cal L}^{\rm NR}_{D_1}=\frac{v_{\phi}^2}{2 \Lambda^2}F^3_{\mu\nu}F^{\mu\nu Y}\,,
\end{equation}
which leads to decays into fermion pairs, into $W^+ W^-$, into $Z \eta$ and
into $Z h$.

For the two-body decay into fermion pairs we obtain (neglecting fermion
masses)
\begin{eqnarray}
  \Gamma(A\rightarrow f \bar{f})
  = \frac{1}{3}\frac{C}{64\pi} \frac{g^2 v_\phi^4}{\Lambda^4}
  \left( (d_V^f)^2 + (d_A^f)^2 \right) M_A\,,
  \label{eqn:WidhtsFermions}
\end{eqnarray}
where we have introduced a color factor which is $C=1$ for leptons and $C=3$
for quarks. The effective vector and axial couplings to neutrinos, charged
leptons, up-type and down-type quarks are given by
\begin{eqnarray}
  d_V^\nu &=& d_A^\nu = -d_A^e = d_A^u = -d_A^d = -\frac{1}{2}\frac{M_A^2}{M_A^2-M_Z^2},\\
  d_V^e   &=& \left( 2 \sin^2\theta_W -\frac{1}{2} \right)\frac{M_Z^2}{M_Z^2-M_A^2}
  -\frac{3}{2},\\
  d_V^u   &=& \left( \frac{1}{2}-\frac{4}{3}\sin^2\theta_W \right)
  \frac{M_Z^2}{M_Z^2-M_A^2} + \frac{5}{6},\\
  d_V^d   &=& \left( \frac{2}{3} \sin^2\theta_W-\frac{1}{2} \right)
  \frac{M_Z^2}{M_Z^2-M_A^2}-\frac{1}{6}\;.
  \label{eqn:DefinitionsD}
\end{eqnarray}
The other decay widths are given by
\bea
\Gamma(A\rightarrow W^+W^-)  &= & \frac{v_\phi^4 }{16 \Lambda^4 } \frac{
\alpha \cos^2\theta_W}{12} M_A \left(\frac{M_A}{M_W}\right)^4
\left(\frac{M_Z^2}{M_Z^2 - M_A^2}\right)^2\nonumber\times\\ &&\times \left(1 +
20 \frac{M_W^2}{M_A^2}+ 12 \frac{M_W^4}{M_A^4}\right) \left(1 - \frac{4
M_W^2}{M_A^2}\right)^{3/2}\;,
\eea
and
\begin{eqnarray}
  \Gamma(A\rightarrow Z \eta)  & = & \frac{1}{3}\frac{1}{256 \pi}\frac{g^2
  M^2_Z \sin^2\beta}{\cos^2\theta_W}\frac{v_{\phi}^4}{\Lambda^4
  M_A}\Big(\frac{M_Z^2}{M_Z^2-M_A^2} -\sin\theta_W\Big)^2 \nonumber\times\\ &
  &\times \Big(10+\frac{M^2_A}{M^2_Z}+\frac{M^2_Z}{M^2_A}+\frac{M^2_{\eta}(M^2_\eta-2
  M^2_A - 2 M^2_Z)}{M^2_A M^2_Z} \Big)\nonumber\times\\ & &\times \sqrt{(1-(M_Z+M_\eta)^2/M^2_A)(1-(M_Z-M_\eta)^2/M^2_A)}\;. 
  \label{eqn:ddr}
\end{eqnarray} 
For the corresponding decay channels into $Z h$, the last equation holds with
the replacement of the physical $h$ boson from Eqs.~(\ref{physscal}).\\

The second term, on the other hand, reads:
\begin{equation}
  {\cal L}^{\rm NR}_{D_2}=\frac{1}{\Lambda^2}
  v_{\phi}\partial_{\mu}A^3_{\nu}(\partial_{\mu}B_{\nu}-\partial_{\nu}B_{\mu})\eta'\,,
\end{equation}
which leads to two body decays involving the hidden sector Higgs $\eta$, using Eqs.~(\ref{physscal}), and a
gauge boson, with rates:
\begin{eqnarray}
  \Gamma(A\rightarrow \gamma\eta) & = & \frac{1}{3} \frac{1}{32\pi}
  \frac{v_{\phi}^2}{\Lambda^4} \cos^2\theta_W \cos^2\beta
  \frac{(M^2_A-M^2_{\eta})^3}{M^3_A} \,,\\ 
  \Gamma(A\rightarrow Z\eta) &
  = & \frac{1}{3} \frac{1}{32\pi} \frac{v_{\phi}^2}{\Lambda^4} \sin^2\theta_W
  \cos^2\beta \times \\&&\times\Big((M^2_A-M^2_{\eta}+M^2_Z)^2+2M^2_Z M^2_A
  \Big)  \frac{\sqrt{\lambda(M_A,M_Z,M_\eta)}}{M_A^3}\;.
  \nonumber
\end{eqnarray}
The decay into $Z h$ and $\gamma h$ can be obtained  with the substitution
$\cos\beta \rightarrow \sin\beta$.

If $\beta\neq 0$ and $\beta\neq \pi/2$, then we have interference between the
decay coming from $\mathcal{L}^{NR}_{D_1}$ and $\mathcal{L}^{NR}_{D_2}$,
which lead to the corrections:
\begin{eqnarray}
\delta\Gamma(A \rightarrow Z \eta) & = &
-\frac{1}{3}\frac{3}{64\pi}\frac{v_{\phi}^3}{\Lambda^4}\frac{g \sin\theta_W
M_Z \sin2\beta}{\cos\theta_W }
\Big(\frac{M_Z^2}{M_Z^2-M_A^2} -\sin\theta_W
\Big)\times\nonumber\\&&\times\Big(M_A^2-M_{\eta}^2+M_Z^2
\Big)\frac{\sqrt{\lambda(M_A,M_Z,M_{\eta})}}{M_A^3}\,,\\
\delta\Gamma(A \rightarrow Z h) & = &
\frac{1}{3}\frac{3}{64\pi}\frac{v_{\phi}^3}{\Lambda^4}\frac{g \sin\theta_W M_Z \sin2\beta}{\cos\theta_W}
\Big(\frac{M_Z^2}{M_Z^2-M_A^2} -\sin\theta_W
\Big)\times\nonumber\\&&\times\Big(M_A^2-M_{h}^2+M_Z^2
\Big)\frac{\sqrt{\lambda(M_A,M_Z,M_{h})}}{M_A^3}\;.
\end{eqnarray}

\section{Cosmic ray signatures of hidden vector dark matter decay}
In this section we firstly give a short introduction to the propagation
of gamma rays and charged cosmic rays through the Galaxy, and secondly
discuss the typical cosmic-ray signatures of hidden vector dark matter,
including strong gamma-ray lines and possible contributions to the anti-matter
fluxes.

\subsection{Cosmic ray propagation}
The hidden gauge boson decay produces a high energy flux of stable particles,
such as gamma rays, electrons, positrons, antiprotons, neutrinos and
antideuterons. The flux of high-energy cosmic rays depends essentially on the
scale of custodial symmetry breaking, which is thus constrained by the
requirement that the predicted fluxes do not exceed the observed fluxes.
We will show that a typical signature of the hidden vector dark matter model
is a prominent gamma-ray line, with values of the custodial symmetry breaking
scale close to the Grand Unification scale and in reach of the Fermi LAT dark matter
searches. Furthermore, as mentioned in the introduction, experiments
measuring the positron fraction and the total electron plus positron flux
indicate the existence of an additional source of electrons and positrons at
high energies, but no additional source of antiprotons. We will also explore
the possibility that the decay of dark matter hidden gauge bosons could be the
origin of such excesses.\\

The production rate of particle $i=e,\gamma,\bar{p}$ per unit energy and unit
volume at a position $\vec{r}$ with respect to the center of the Milky Way is
given by
\begin{equation}
  Q_i(E,\vec{r})=\frac{\rho(\vec{r})}{M_{\rm DM}\,\tau_{\rm DM}}\frac{dN_i}{dE}\;,
  \label{source-term}
\end{equation}
where $dN_i/dE$ is the corresponding energy spectrum of particle $i$ produced
in the decay, and $\rho(\vec{r})$ is the density profile of dark matter
particles in the Milky Way halo. We will adopt in this paper the spherically
symmetric NFW density profile~\cite{NFW96} for definiteness,
\begin{equation}
  \rho(r)\propto \frac{1}{(r/r_c) [1+(r/r_c)]^2}\;,
  \label{eqn:NFW}
\end{equation}
normalized to $0.3 \GeV/\cm^3$ at the position of the sun, $r=8.5\kpc$,
although our results do not depend much on the specific form of the halo
profile.\\

Gamma-rays, contrary to electrons, positrons and antiprotons, which will be
discussed below, do not diffuse and carry information about their spatial
origin. The gamma-ray signal from dark matter decay consists of several components. The
most important one is related to the prompt radiation (\textit{e.g.}~final
state radiation) produced in the decay of DM particles inside the Milky Way
halo. It depends on the halo density profile, and although the halo profile is
expected to be approximately isotropic, the corresponding flux at Earth
exhibits a dipole-like anisotropy which is due to the off-set between 
the Sun and the Galactic center and which can be as large as 20-30\% 
for dark matter lifetimes of the order of $10^{26}$s~\cite{Ibarra:2009nw}.
In contrast, the
extragalactic prompt component of the $\gamma$-ray signal, which stems from
the decay of dark matter particles at cosmological distances, is isotropic. At energies
around $10\,\text{GeV}$ or below, the magnitude of the halo and extragalactic
fluxes are comparable when looking in direction of the anti-galactic center,
whereas at higher energies around and above $100\,\text{GeV}$ the inelastic
scattering between $\gamma$-rays and the intergalactic background light
reduces the extragalactic component considerably (see
Ref.~\cite{Ibarra:2009nw} for a discussion). In the plots we assume a $10\%$
energy resolution, and we show also the H.E.S.S~\cite{A08a, Aharonian:2009ah}
results for the electron + positron (+ gamma) flux in the TeV region, which
acts like an upper bound on the isotropic flux.

For details about the adopted calculation of the electron, anti-matter and
gamma-ray fluxes we refer to Refs.~\cite{Ibarra:2009dr, Ibarra:2009nw}.

Electrons and positrons from dark matter decay loose their energy mainly via
interaction with the Galactic magnetic field and the interstellar radiation
field (ISRF).  In the first case (assuming injection energies of the order of
$1\TeV$) synchrotron radiation in the radio band with frequencies
$\mathcal{O}(0.1 - 100~\text{GHz})$ is produced and potentially observable
(see {\it e.g.} Refs.~\cite{Ishiwata:2008qy, Zhang:2008tb, Zhang:2009pr}). In
the second case, the inverse Compton scattering (ICS) of electrons and
positrons on the ISRF (which includes the cosmic microwave background, thermal
dust radiation and starlight) produces a second component of gamma rays with
energies between $100\MeV$ and $1\TeV$~\cite{Cirelli:2009vg, Ishiwata:2009dk}.
However, since electron and positron fluxes are always relatively weak in the
decay channels we consider we will neglect ICS radiation throughout this work
for simplicity.\\

After being produced in the decay of dark matter particles, electrons and
positrons scatter on irregularities of the Galactic magnetic field before
reaching the Earth, which results in a wash-out of directional information.
Their propagation is commonly described by a diffusion model, whose free
parameters are tuned to reproduce the observed cosmic-ray nuclei
fluxes~\cite{GDB+}. As propagation parameters we will adopt the ones of
the MED propagation model defined in \cite{MDT+01}, which provide the best fit
to the Boron-to-Carbon (B/C) ratio: $\delta=0.70$, $K_0=0.0112\,{\rm
kpc}^2/{\rm Myr}$ and $L=4\,{\rm kpc}$. Our conclusions, however, are rather
insensitive to the choice of propagation parameters. The astrophysical
background in the $e^\pm$-channel is mainly due to primary electrons, which
are presumably produced in supernova remnants, and due to secondary positrons,
produced in the interaction of cosmic-rays with the galactic gas. For these
background fluxes we adopt the ``Model 0'' presented by the Fermi
collaboration in \cite{Grasso:2009ma}, which fits well the low-energy data
points of the total electron plus positron and the positron fraction.
We allow, however, for a 10\% rescaling of the electron background in order to improve the agreement of the total flux to the data.

The antiproton propagation in the Galaxy is analogous to the propagation of
electrons and positrons. However, since antiprotons are much heavier than
electrons and positrons, energy losses are negligible. Furthermore, antiproton
propagation is affected by convection, which accounts for the drift of
antiprotons away from the disk induced by the Milky Way's Galactic wind. For
predictions of the antiproton flux we show an error band, corresponding to the
MIN and MAX model of Ref.~\cite{MDT+01}. In our plots we present actually the
$\bar{p}/p$-ratio, where we adopt the proton and anti-proton backgrounds from
Ref.~\cite{Lionetto:2005jd}.

For both, electrons/positrons and anti-protons, the fluxes at the top of the
atmosphere can differ considerably from the interstellar fluxes at energies
smaller than $\sim 10$ GeV, due to solar modulation effects. To take this
effect into account, we adopt the force field approximation~\cite{GA67} with
$\phi_F=550$ MV~\cite{Bar97}.\\

The main background in the $\gamma$-ray channel is the diffuse emission of our
Galaxy, which is mainly due to interactions of cosmic rays with the galactic
gas and the ISRF~\cite{Strong:1998fr}. This component is by far strongest in
the galactic disk region, and it turns out that exotic fluxes from dark matter decay
would dominantly show up at higher latitudes, away from the disk. For this
reason they could be misidentified as contribution to the extragalactic
gamma-ray background (although they can be distinguished by their large scale
anisotropy, see Ref.~\cite{Ibarra:2009nw}). In this work we will show
predictions for the averaged gamma-ray flux in the region 
$0\leq l \leq 360^\circ$, $10^\circ\leq |b| \leq 90^\circ$, which offers
the best strategy for searching gamma-ray lines from dark matter 
decay~\cite{Bertone:2007aw}.

\subsection{Gamma-ray lines}
The existence of two-body decay modes with gamma-ray lines in the final state
are a generic prediction of hidden vector dark matter models. We will discuss
this for each possible operator separately, Eqs.~\eqref{eqn:opA}-\eqref{eqn:opD}.

\begin{table}[t]
  \begin{center}
    \begin{tabular}{c|cccccc}
      \hline\hline
      Benchmark & $M_A$ & $g_\phi$ & $v_\phi$ & $M_\eta$ & $M_h$ &$\sin \beta$ \\
      \hline\hline
      1 & 300 GeV & 0.55   & 1090 GeV & 30 GeV  & 150 GeV & $\approx0$ \\
      2 & 600 GeV &  0.6   & 2000 GeV & 30 GeV  & 120 GeV & $\approx0$ \\
      3 &  14 TeV &  12    & 2333 GeV & 500 GeV & 145 GeV & $\approx0$ \\
      4 &1550 GeV &  2.1   & 1457 GeV & 1245 GeV & 153 GeV & 0.25 \\
      \hline\hline
    \end{tabular}
    \caption{Benchmark points used for the calculation of cosmic-ray
    signatures.}
    \label{tab:bmp}
  \end{center}
\end{table}

\paragraph{Case A and B.}
In cases A and B, Eqs.~\eqref{eqn:opA} and~\eqref{eqn:opB}, the dark matter
particle decays either into two scalar particles ($\eta\eta$, $h \eta$, $h h$)
or into a gauge boson and a scalar particle  ($\gamma\eta$, $\gamma h$,
$Z\eta$, $Z h$).  Whether the dark matter particle decays preferentially into
two scalar particles or into a gauge boson and a scalar particle depends on
the details of the model. In both cases, the fragmentation and decay of the
Higgs boson or the hidden sector $\eta$ boson could produce a sizable flux of
electrons, positrons and antiprotons.  Unfortunately, the electrons and
positrons produced in fragmentations cannot explain the PAMELA and Fermi
excesses and moreover the PAMELA measurements on the antiproton-to-proton
ratio set very stringent constraints on possible new sources of antiprotons.
Interestingly, even if the scale $\Lambda$ is increased in order to avoid an
antiproton excess, the generically present gamma-ray lines can still be
intense enough to be observed in experiments,  due to the enormous sensitivity
of dark matter line searches.

\begin{figure}[t]
  \begin{center}
    \includegraphics[width=\picwid\linewidth]{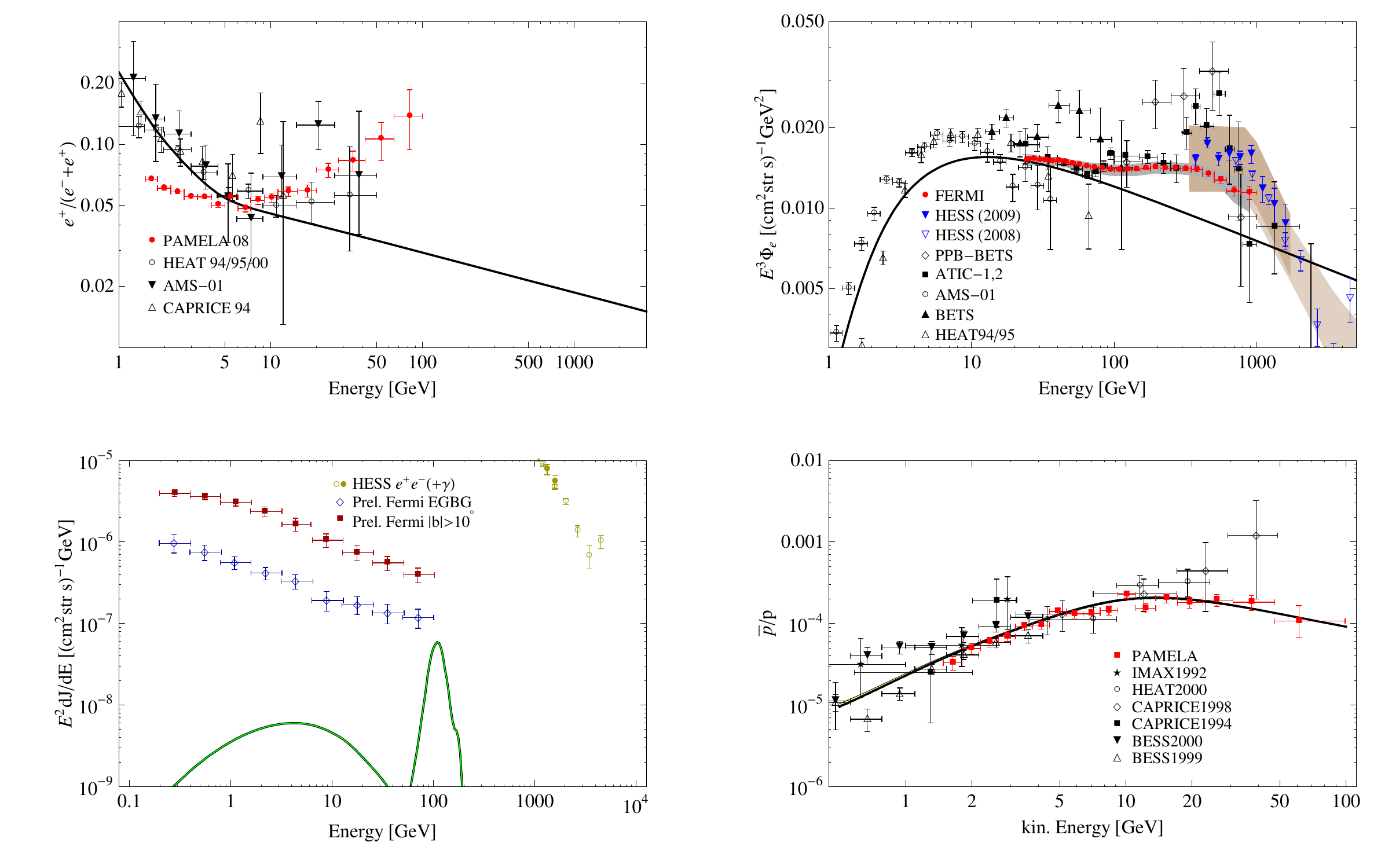}
  \end{center}
  \vspace{-0.5cm}
  \caption{Predictions for case A, benchmark 1, with
  $\tau_\text{DM}=1.7\times10^{28}\s$ ($\Lambda = 2.9\times10^{15}\GeV$). The
  \textit{upper panels} show the positron fraction (\textit{left}) and the
  total electron + positron flux (\textit{right}) compared with experimental
  data. \textit{Dashed} lines show the adopted astrophysical background,
  \textit{solid} lines are background + dark matter signal (which overlap the
  background in this plot). The \textit{lower left panel} shows the gamma-ray
  signal from dark matter decay, whereas the \textit{lower right panel} shows
  the $\bar{p}/p$-ratio: background (\textit{dashed line}) and overall flux
  (\textit{solid lines}, again identical with background). }
  \label{fig:A5}
\end{figure}

\begin{figure}[t]
  \begin{center}
    \includegraphics[width=\picwid\linewidth]{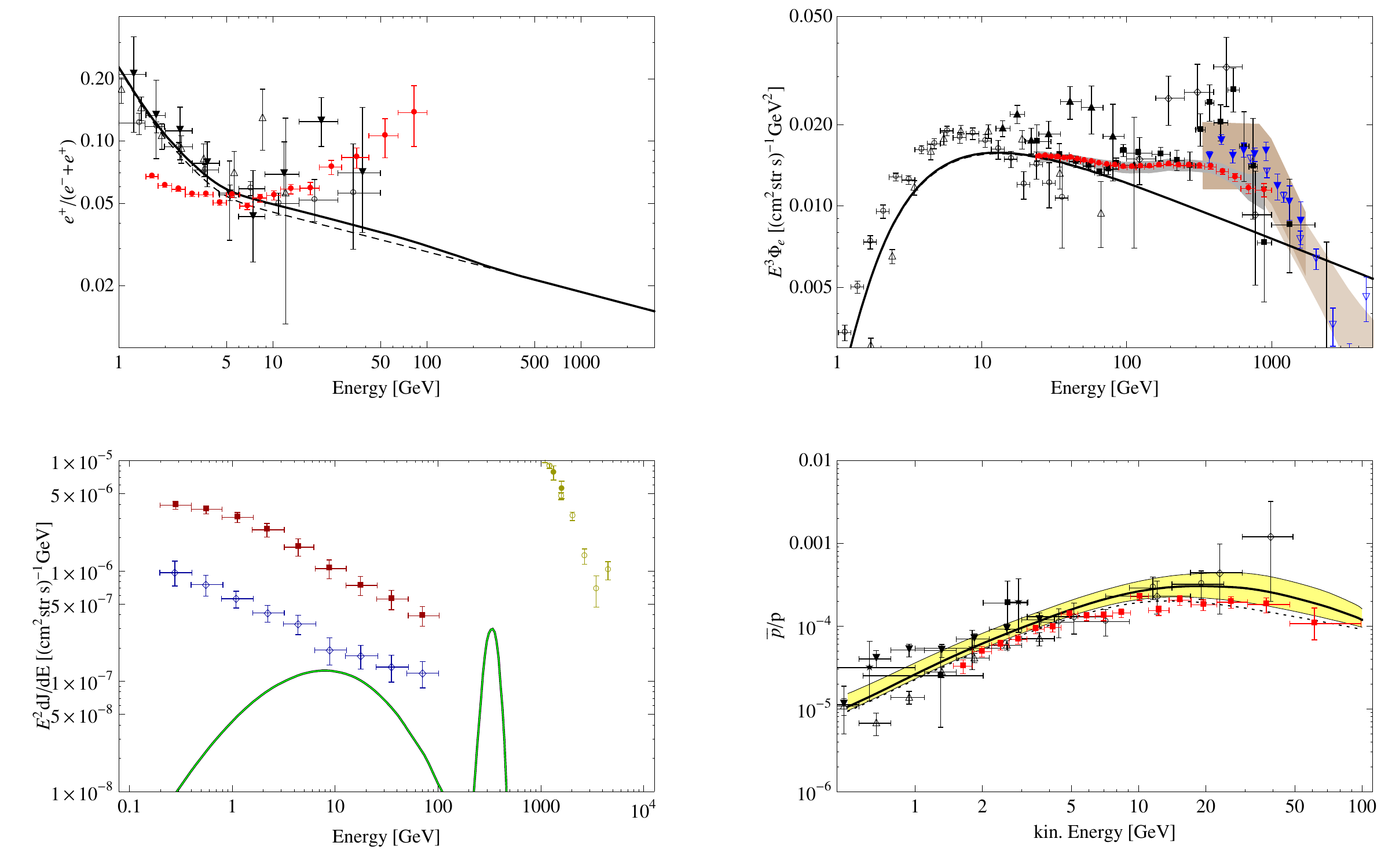}
  \end{center}
  \vspace{-0.5cm}
  \caption{Like Fig.~\ref{fig:A5}, but for case A, benchmark 2, with
  $\tau_\text{DM}=1.1\times10^{27}\s$ ($\Lambda = 3.7\times10^{15}\GeV$). The
  yellow band shows the uncertainty in the anti-proton flux due to the
  propagation model parameters.}
  \label{fig:A10}
\end{figure}

This is illustrated in Figs.~\ref{fig:A5} and \ref{fig:A10}, where we show 
the predictions for the positron fraction, total electron 
plus positron flux, antiproton-to-proton fraction and gamma-ray flux
for two generic scenarios, namely the benchmark points 1 and 2 defined in 
Tab.~\ref{tab:bmp}. These
choices of parameters can successfully reproduce the observed dark matter
abundance and are consistent with all present laboratory constraints.  
We also show in the plots  for the 
positron fraction the results from PAMELA~\cite{A08}, HEAT~\cite{B97},
CAPRICE~\cite{BCF+00} and AMS-01~\cite{A07};  for the total
electron plus positron flux, the results from
Fermi~\cite{Abdo:2009zk}, H.E.S.S.~\cite{A08a, Aharonian:2009ah}, BETS,
PPB-BETS~\cite{T08}, ATIC~\cite{C08a}, HEAT, CAPRICE and AMS-01; 
for the antiproton-to-proton ratio, the results from PAMELA~\cite{A09}, 
BESS95~\cite{M98}, BESS95/97~\cite{O00}, CAPRICE94~\cite{BCF+97}, 
CAPRICE98~\cite{BBS+01} and IMAX~\cite{M96} and for the gamma-ray flux,
the preliminary data from the Fermi-LAT in the region between $10^\circ$ and 
$90^\circ$, as well as the extraction of the extragalactic flux from these
data~\cite{talkAckermann}. In the gamma-ray plot, we also show the H.E.S.S.
results for the electron + positron (+gamma) flux at high energies, which
yields also an upper bound on the overall isotropic gamma-ray flux.

The branching ratios for these two benchmark points are listed in
Tab.~\ref{tab:BRA}. Benchmark point 1 is characterized by large branching
ratios into gauge boson and Higgs, being the decay into a monoenergetic gamma
line the dominant channel. On the other hand, since kinematically allowed,
benchmark point 2 is characterized by a large branching ratio into two Higgs
bosons. It is interesting that, even though the decay mode into monoenergetic
gamma rays is subdominant in this benchmark point, the gamma-ray line still is
a very prominent feature in the gamma-ray energy spectrum. 

We estimate that, in the limit $v_\phi \gg v$, $\beta\rightarrow 0$,
the decay rate into $\gamma h$ is given by:
\begin{equation}
  \Gamma(A\rightarrow \gamma h)^{-1} = 1.5\times10^{28}\s
  \left( \frac{\Lambda}{2\times 10^{15}\GeV} \right)^4
  \left( \frac{1\TeV}{v_\phi} \right)^2 
  \left( \frac{100 \GeV}{M_A} \right)\;.
  \label{eqn:}
\end{equation}
The Fermi-LAT observations of the region $|b| > 10^\circ$ 
plus a $20^\circ \times 20^\circ$ square around the Galactic center constrain
the dark matter lifetime to be longer than a few times $10^{28}\s$
at energies below 200 GeV~\cite{Abdo:2010nc}, which is taken into account in the
results shown for benchmark point 1 (see Fig.~\ref{fig:A5}), where the line is
around 110 GeV. Thus present experiments can probe values of the scale of
custodial symmetry breaking close to the Grand Unification scale. In case of
benchmark point 2 the line occurs at an energy scale above the ones probed by Fermi, 
so that smaller lifetime are allowed experimentally.  We
show results for a lifetime $1.1\times 10^{27}$~s, where the contributions to the diffuse
gamma-rays around 10 GeV and the anti-proton fluxes can be sizeable. The gamma line in this case is huge and should be seen by any experiment sensitive to these energies.

\begin{table}[t]
  \centering
  \begin{tabular}{c|ccccccc}
    \hline\hline
    Benchmark & $\eta\eta$ & $h\eta$ & $hh$ & $\gamma\eta$ & $Z\eta$ & $\gamma
    h$ &
    $Zh$\\\hline\hline
    1 & - & 0.09& -    & 0.04 & 0.02 & 0.65 & 0.20 \\
    2 & - & 0.04& 0.62 & 0.002 & 0.003 & 0.15 & 0.18 \\
    3 & - & 0.04 & 0.80 & $3\times10^{-6}$ & 0.002 & 0.0003 & 0.16 \\\hline\hline
  \end{tabular}
  \caption{Branching Ratios for Case A.}
  \label{tab:BRA}
\end{table}


\begin{figure}[t]
  \begin{center}
    \includegraphics[width=\picwid\linewidth]{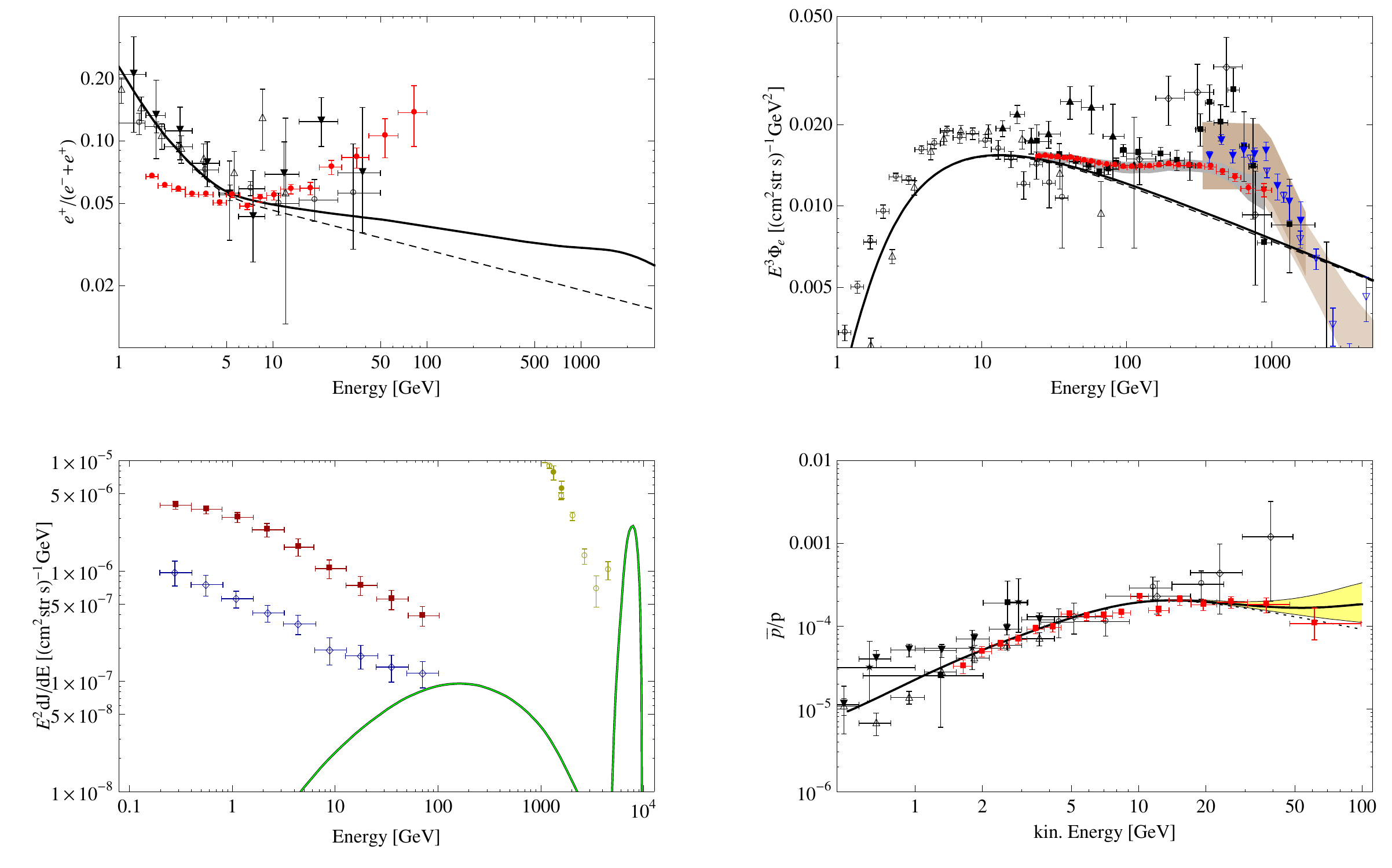}
  \end{center}
  \vspace{-0.5cm}
  \caption{Like Fig.~\ref{fig:A5}, but for case C, benchmark 3, with
  $\tau_\text{DM}=6.0\times10^{26}\s$ ($\Lambda = 2.0\times10^{17}\GeV$).}
  \label{fig:C9}
\end{figure}

\paragraph{Case C.}
This operator, see Eq.~\eqref{eqn:opC}, predicts decays only into a gauge
boson and a scalar particle, either $h$ or $\eta$. 
A large decay branching ratio into monoenergetic gamma-rays is predicted unavoidably, as
illustrated in Tab.~\ref{tab:BRC} for the four different benchmark scenarios.
In the limit $M_\eta\ll M_A$, the decay rate into $\gamma \eta$ is given by
\begin{equation}
  \Gamma(A\rightarrow \gamma\eta)^{-1} = 2.7\times10^{28}\s
  \left( \frac{\Lambda}{4\times10^{15}\GeV} \right)^4
  \left( \frac{300 \GeV}{M_A} \right)^5\;.
  \label{eqn:}
\end{equation}
which can make the gamma-ray line observable at the Fermi-LAT for values of
the scale of custodial symmetry close to the Grand Unification Scale,
especially for large dark matter masses. The cosmic ray signatures of
benchmark point 1 for case C are very similar to case A, {\it cf.}
Fig.~\ref{fig:A5}. On the other hand, we show in Fig.~\ref{fig:C9} the
predictions for benchmark point 3 with a very large dark matter mass of 14
TeV, which predicts a strong line at very high energies and only small
contributions to positrons and anti-protons.

\begin{figure}[t]
  \begin{center}
    \includegraphics[width=\picwid\linewidth]{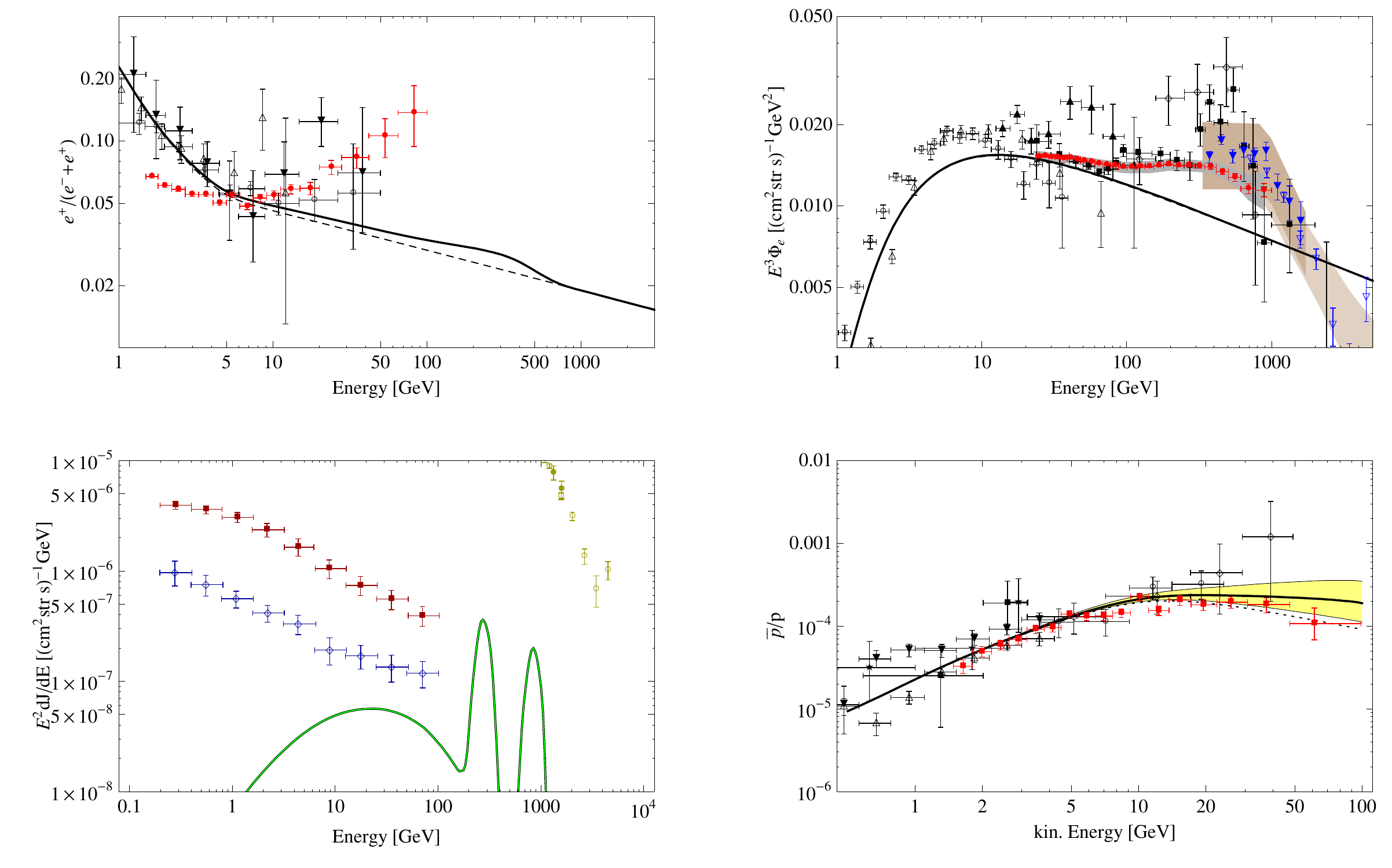}
  \end{center}
  \vspace{-0.5cm}
  \caption{Like Fig.~\ref{fig:A5}, but for case C, benchmark 4, with
  $\tau_\text{DM}=1.6\times10^{27}\s$ ($\Lambda = 1.2\times10^{16}\GeV$).}
  \label{fig:Cf}
\end{figure}

One feature of the model that is in principle present for each
operator, and which we want to illustrate for case C, is the general existence
of two independent gamma-ray lines. These lines stem from the decay into
$\gamma h$ and $\gamma\eta$ and would appear at different energies as long as
the higgs and the $\eta$ masses are not too degenerate. In case C both of the
decay channels are in general open as long as $\sin\beta\neq0$, which is the
case for benchmark point 4 in Tab.~\ref{tab:bmp}. In Fig.~\ref{fig:Cf} we show
the corresponding cosmic-ray fluxes. Most interestingly the gamma-ray flux
exhibits two strong peaks in this case, located at 270 and 770 GeV.

\begin{table}[t]
  \centering
  \begin{tabular}{c|cccc}
    \hline\hline
    Benchmark & $Z\eta$ & $\gamma\eta$ & $Zh$ & $\gamma h$ \\\hline\hline
    1 & 0.19 & 0.81 & 0 & 0 \\
    2 & 0.22 & 0.78 & 0 & 0 \\
    3 & 0.23 & 0.77 & 0 & 0 \\
    4 & 0.028 & 0.79 & 0.041 & 0.14 \\
    \hline\hline
  \end{tabular}
  \caption{Branching Ratios for Case C, including benchmark point 4 which
  features decay channels with $h$ in the final state.}
  \label{tab:BRC}
\end{table}

\paragraph{Case D.}
This operator, see Eq.~\eqref{eqn:opD}, is particularly interesting since it
induces a kinetic mixing between the $U(1)_Y$ of hypercharge and one of the
hidden $SU(2)$ gauge bosons.  As a result two-body decay modes into lepton and
quark pairs are allowed, in contrast to the other operators. This leads to
interesting implications for the electron/positron flux that will be discussed
shortly below.

Here we firstly emphasize that again the operator also predicts two-body decay
into $\gamma h$, which could be observable in different parts of the parameter
space. The inverse decay rate reads, for $M_\eta \ll M_A$:
\begin{equation}
  \Gamma(A\rightarrow \gamma\eta)^{-1} = 2.4\times10^{28}\s
  \left( \frac{\Lambda}{7\times10^{15}\GeV}\right)^4
  \left( \frac{1\TeV}{v_\phi} \right)^2 
  \left( \frac{300 \GeV}{M_A} \right)^3\;,
  \label{eqn:}
\end{equation}
and shows that the line could be observed by Fermi LAT for scales of the
custodial symmetry breaking close to the Grand Unification scale. For these large lifetimes
around $10^{28}\,$s contributions to the anti-matter channel would be
negligible. However, if the line lies above around 300 GeV and out of reach of
Fermi LAT, shorter lifetimes cannot be excluded and the anti-matter fluxes can
be sizeable.

\subsection{Positron flux}

\begin{table}[t]
  \centering
  \begin{tabular}{c|cccccccc}
    \hline\hline
    Benchmark & $Z\eta$ & $Zh$ & $\gamma\eta$ & $W^+W^-$ & $\nu\bar{\nu}$ & $e^+e^-$ &
    $u\bar{u}$ & $d\bar{d} $ \\\hline\hline
    1 & 0.01 & 0.005 & 0.04 & 0.02 & 0.09 & 0.39 & 0.29 & 0.15\\
    2 & 0.019 & 0.004 & 0.036 & 0.014 & 0.072 & 0.35 & 0.39 & 0.12 \\
    3 & 0.22 & 0.0002 & 0.73 & 0.0005 & 0.003 & 0.016 & 0.018 & 0.005 \\\hline\hline
  \end{tabular}
  \caption{Branching Ratios for Case D.}
  \label{tab:BRD}
\end{table}

Here we will briefly discuss the predictions for the anti-matter fluxes
concentrating on case D, since this operator features two-body decay into
fermions pairs due to effective kinetic mixing between hidden sector and the
hypercharge $U(1)_Y$.  The corresponding branching ratios are listed in
Tab.~\ref{tab:BRD}. In the cases with lower dark matter mass, the branching
ratio into hard leptons (and in particular electrons) is sizable. Namely, in
the limit $M_A\gg M_Z$ the inverse decay rate into charged lepton pairs is
given by
\begin{equation}
  \Gamma(A\rightarrow \ell^+ \ell^-)^{-1} = 2.6\times10^{27}\s
  \left( \frac{\Lambda}{7\times10^{15}\GeV} \right)^4
  \left( \frac{1\TeV}{v_\phi} \right)^4 
  \left( \frac{300 \GeV}{M_A} \right)\;,
  \label{eqn:}
\end{equation}
which can produce a steep rise in the observed positron fraction for values of
the scale of custodial symmetry breaking of the order of the Grand Unification
scale. 

\begin{figure}[t]
  \begin{center}
    \includegraphics[width=\picwid\linewidth]{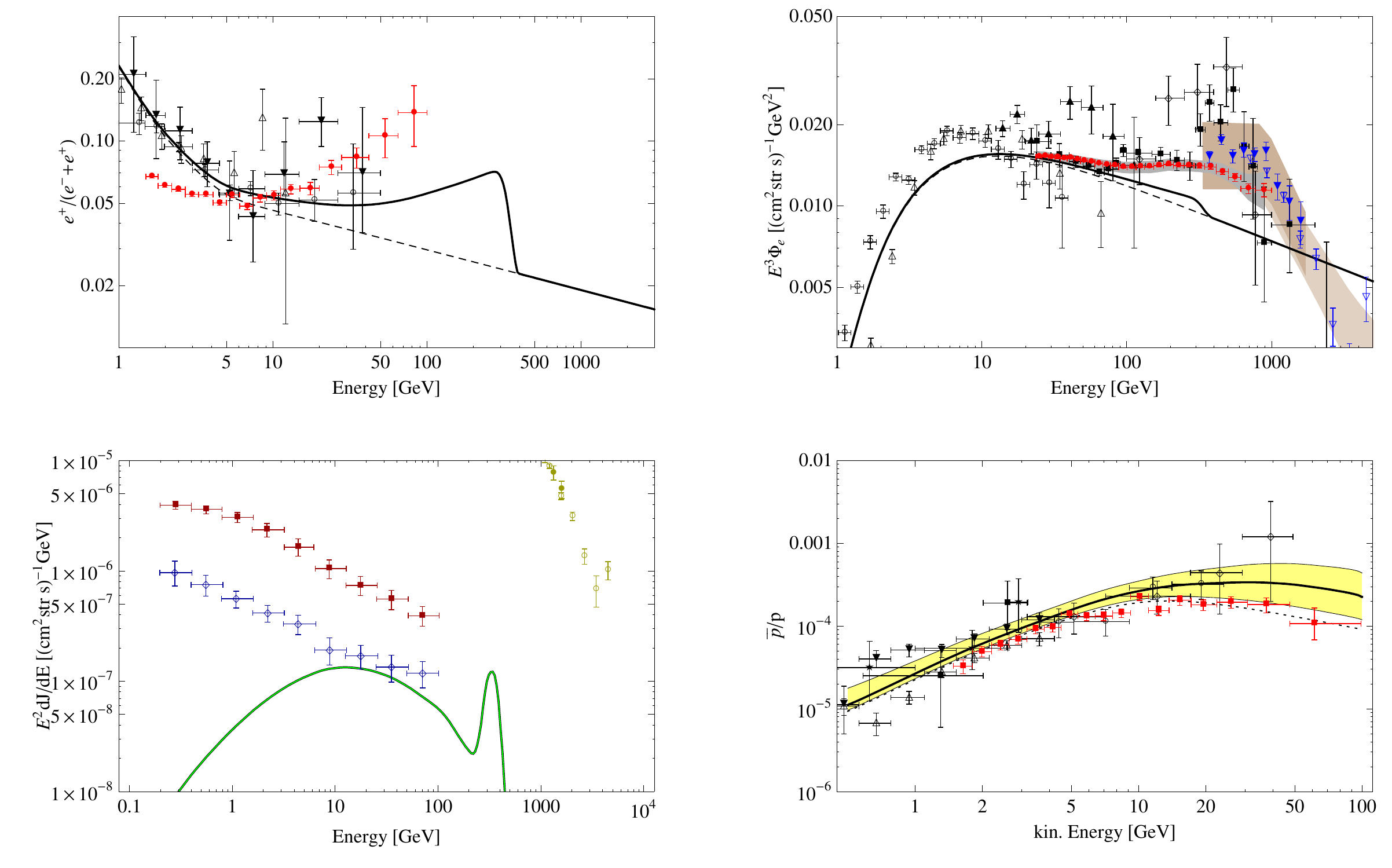}
  \end{center}
  \vspace{-0.5cm}
  \caption{Like Fig.~\ref{fig:A5}, but for case D, benchmark 2, with
  $\tau_\text{DM}=6.7\times10^{26}\s$ ($\Lambda = 1.5\times10^{16}\GeV$).}
  \label{fig:D10}
\end{figure}

As an example, we show in Fig.~\ref{fig:D10} the predictions for the cosmic
ray fluxes for benchmark point 2. For a scale of custodial symmetry breaking
$\Lambda=7.2\times 10^{15}\GeV$, which is close to the Grand Unification scale,
the gamma
ray spectrum shows a intense gamma-ray line at 300 GeV, in agreement with
current observations. On the other hand, the positron fraction shows a steep
rise which could partially, although not totally, contribute to the PAMELA
positron excess. Moreover, the decay into charged leptons is necessarily
accompanied by a decay into quarks, which produce a sizable antiproton flux
and is in some tension with the observations. This is a generic feature of the
decay mode and hence it is unlikely that it contributes the dominant part to
the observed positron excess. 

In more generality we found that the PAMELA and Fermi results
can be reproduced in principle by the model, but only at the price of
producing a too large diffuse $\gamma$ signal, too many antiprotons (unless the dark matter is very heavy) and
sometimes gamma lines above the rates allowed by the H.E.S.S. measurements in
the multi TeV range.

\paragraph{Discussion.} 
It is intriguing that the production of a $\gamma$-ray line is a generic
prediction for all possible operators that may mediate the decay of the
$SU(2)_{\rm HS}$ dark matter gauge bosons. For values of the custodial symmetry
breaking scale near to the Grand Unification scale, and for dark matter masses around
$400\GeV$ and below, this line could be in reach of sensitivity of the Fermi LAT gamma-ray line searches. On the other hand, a production of an
observable amount of electrons and positrons or anti-protons is very model
dependent. In most cases electrons and positrons are produced in the
fragmentation of scalar or vector bosons and lead to a very flat spectrum. An
interesting exception occurs for the operator case D which features two-body
decay modes into lepton pairs. In this case the produced positron spectrum can
rise more steeply, but, when also taking other observations into account,
still not enough to explain the PAMELA observations alone.

\section{Effects of the annihilation processes with one dark matter particle in the
final state}

The model considered above has the interesting and rather peculiar property
that it allows annihilation processes with one dark matter particle in the final state,
i.e.~$A_i A_j \rightarrow A_k \eta$ annihilations via an intermediate $A_k$,
Fig.~2 of \cite{Hambye:2008bq}.  In ordinary models based on a $Z_2$ symmetry
such processes are strictly forbidden, they would be equivalent to $Z_2$
breaking at the renormalizable level and therefore to fast DM decay. The
non-abelian character of the custodial symmetry responsible for the stability
of the hidden vectors allows these processes through the trilinear coupling
${\cal L}\owns- \frac{1}{4}  F^{\mu\nu}F_{\mu \nu} \owns -\frac{1}{2} g_\phi
\varepsilon_{ijk} A_j^{\mu} A_k^{\nu} (\partial_\mu A_{i\nu}-\partial_\nu
A_{i\mu})$.  As pointed out in Refs.\cite{Hambye:2008bq,Hambye:2009fg} these
``trilinear" processes do not bring nevertheless any new radical change in the
freeze-out mechanism. In the Boltzmann equations (where  $n = n_1 + n_2 + n_3 $
is the density of $A$ states) 
\begin{equation}
{d n\over dt} + 3 H n = - {\langle
\sigma_{ii} v\rangle\over 3} \left(n^2 - n_{Eq}^2\right) - {\langle\sigma_{ij}
v\rangle\over 3} n (n - n_{Eq})\,,
\end{equation}
these terms, parametrized by
$\sigma_{ij}$,  behave linearly in $n-n_{Eq}$, whereas the ordinary
annihilations, parametrized by $\sigma_{ii}$, behave linearly in
$n^2-n^2_{Eq}$.  Since $n^2 - n_{Eq}^2 \approx 2 n(n-n_{EQ})$ near freeze-out,
the relic abundance behaves as usual $\Omega_{DM} \propto 1/\mbox{\rm
Max}(\sigma_{ij}, 2 \sigma_{ii})$.  However these ``trilinear" processes
contribute with a rate expected to be similar to the one of the processes with
no dark matter particle in the final states and consequently should be properly taken
into account. This is what is done here, for ``small" Higgs portal coupling,
$\lambda_m < 10^{-3}$, and for ``large" Higgs portal coupling, $\lambda_m >
10^{-3}$. 

\begin{figure}[t!]
  \begin{minipage}[t]{0.33\textwidth}
    \centering
    \includegraphics[width=0.985\columnwidth]{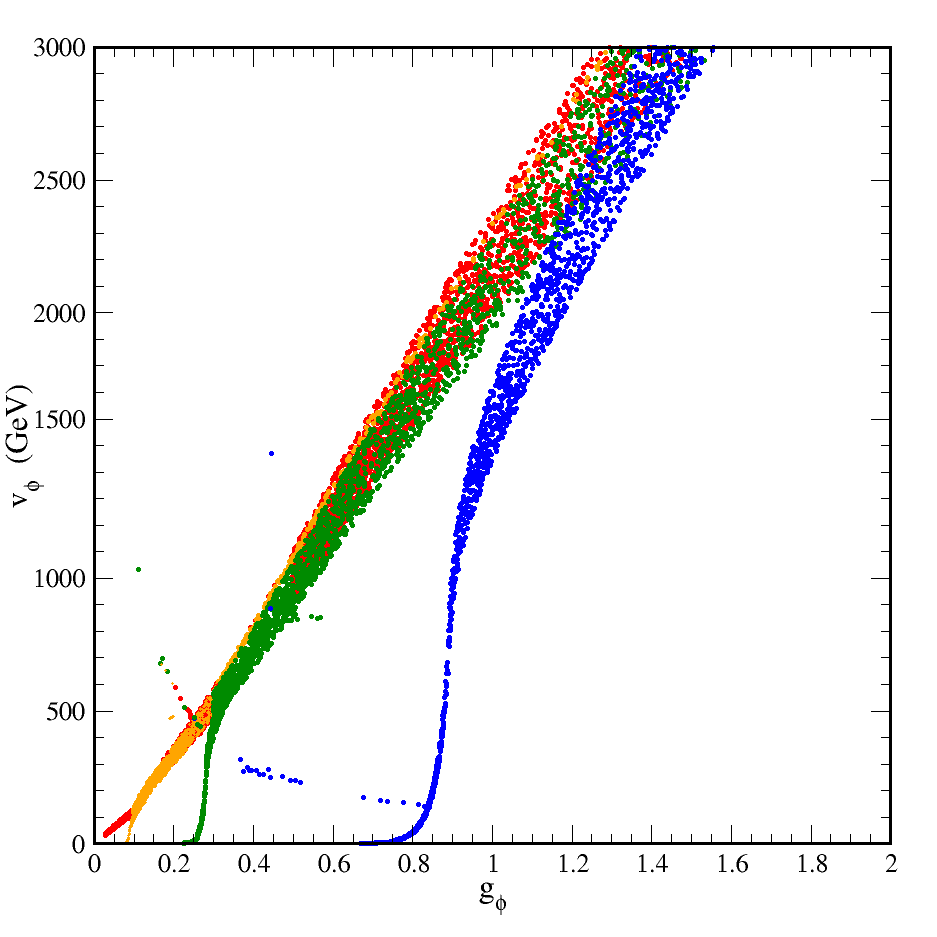}
  \end{minipage}
  \hspace*{-0.2cm}
  \begin{minipage}[t]{0.33\textwidth}
    \centering
    \includegraphics[width=0.985\columnwidth]{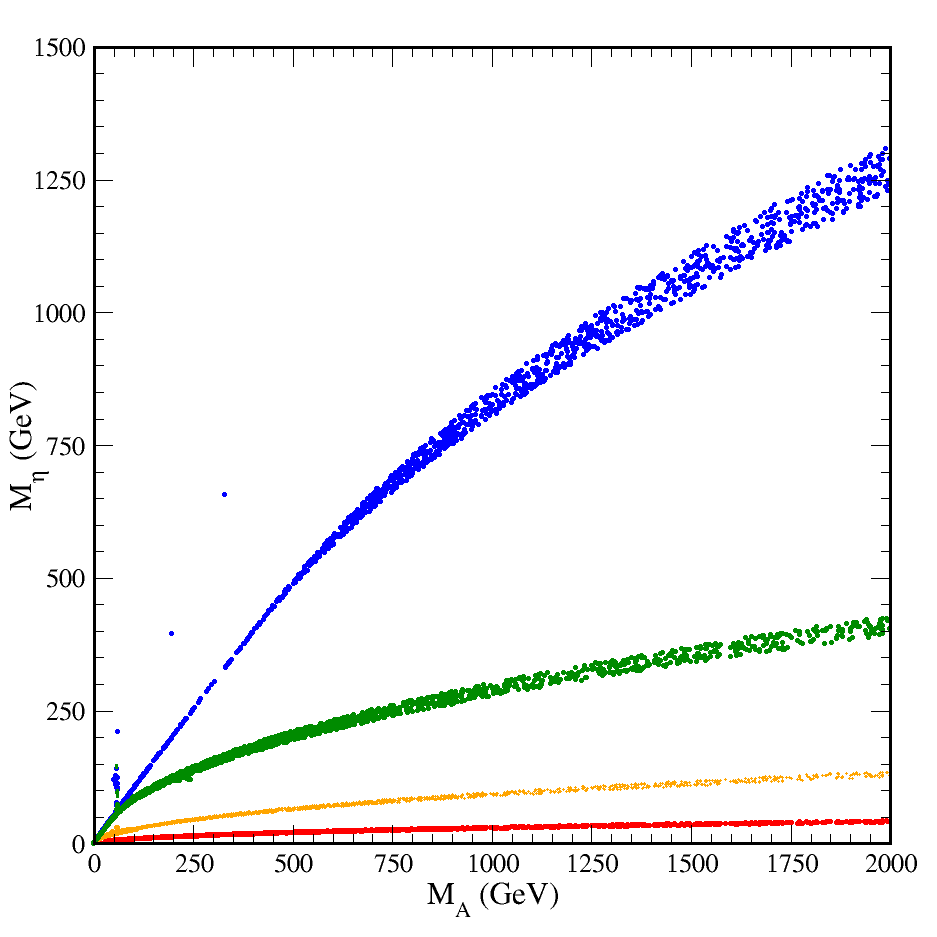}
  \end{minipage}
  \hspace*{-0.2cm}
  \begin{minipage}[t]{0.33\textwidth}
    \centering
    \includegraphics[width=0.985\columnwidth]{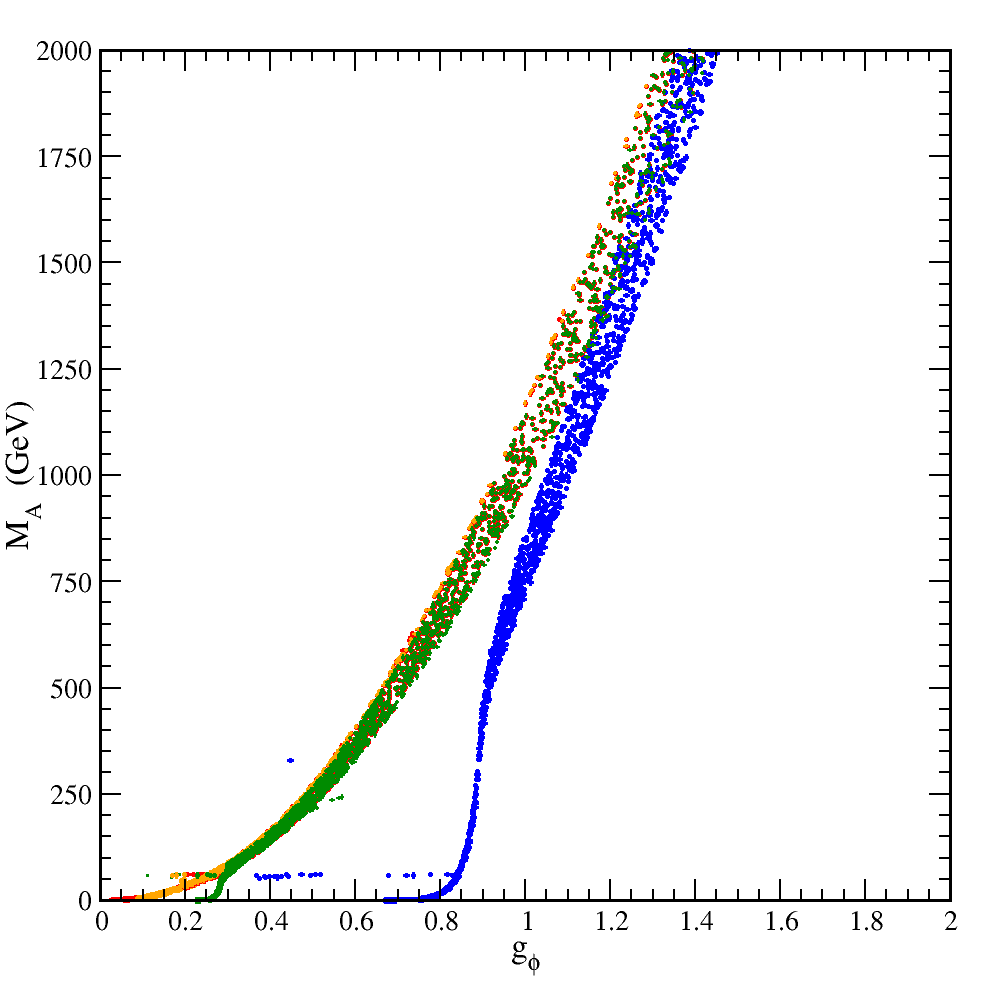}
  \end{minipage}
  \caption{Parameter space leading to $0.091 <\Omega h^2 < 0.129$ for the
  small $\lambda_m$ regime ($10^{-7} < \lambda_m < 10^{-3}$). From left to
  right: $v_{\phi}$ vs $g_{\phi}$, $M_{\eta}$ vs $M_A$ and $M_A$
  vs $g_{\phi}$. The different curves are for various values of
  $\lambda_{\phi}$: $\lambda_{\phi}=10^{-4}$ (red), $\lambda_{\phi}=10^{-3}$
  (orange), $\lambda_{\phi}=10^{-2}$ (green) and $\lambda_{\phi}=10^{-1}$
  (blue). The Higgs mass is fixed at $M_h= 120$ GeV. The dots off the main ``sequences" correspond to Higgs or $\eta$ resonant annihilations, 
 for $M_A
= g_{\phi}v_{\phi}/2 \sim M_h/2$ and $M_{\eta}/2$ respectively.}
\label{fig:slm_ps}
\end{figure}

Considering first the small $\lambda_m$ regime, in Fig~\ref{fig:slm_ps} are
shown the values of the gauge coupling $g_{\phi}$ vs $v_{\phi}$, $M_A$ vs
$M_{\eta}$ and $M_A$ vs $g_{\phi}$,
which lead to a relic density
within the WMAP range $0.091< \Omega h^2 < 0.129$ at
3$\sigma$~\cite{Komatsu:2008hk}, in agreement with the direct detection
experimental upper bounds from CDMS~\cite{Ahmed:2008eu} and
Xenon10~\cite{Angle:2008we} (see Fig.~\ref{fig:dirdet} below). These graphs show
corrections of order unity with respect to the corresponding result without
the ``trilinear" processes, Fig.~3 of Ref.~\cite{Hambye:2008bq}. 
The dominant
processes are the annihilations $A_i A_i \rightarrow \eta \eta$ and $A_i
A_j \rightarrow \eta A_{k}$ which,
unless 
the $\lambda_\phi$ coupling is large, 
have cross-sections proportional to
$g^4_{\phi}/M^2_A$, leading to a 
$M_A \propto g_{\phi}^2$ quadratic behavior in Fig.~\ref{fig:slm_ps}.
The only exception to this behavior is given by the resonant cases, 
when $M_A \sim M_{h}/2$ or $M_A \sim M_{\eta}/2$. 

\begin{figure}[t!]
  \begin{minipage}[t]{0.33\textwidth}
    \centering
    \includegraphics[width=0.985\columnwidth]{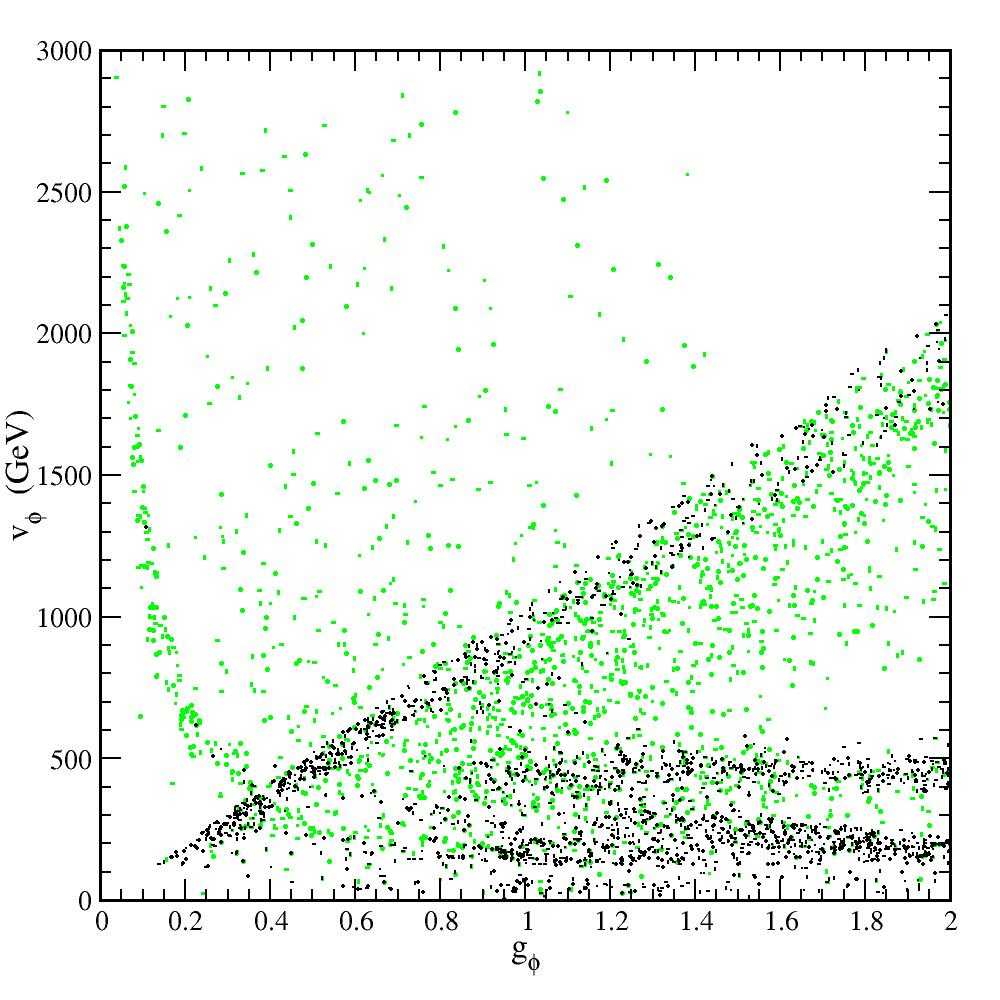}
  \end{minipage}
  \hspace*{-0.2cm}
  \begin{minipage}[t]{0.33\textwidth}
    \centering
    \includegraphics[width=0.985\columnwidth]{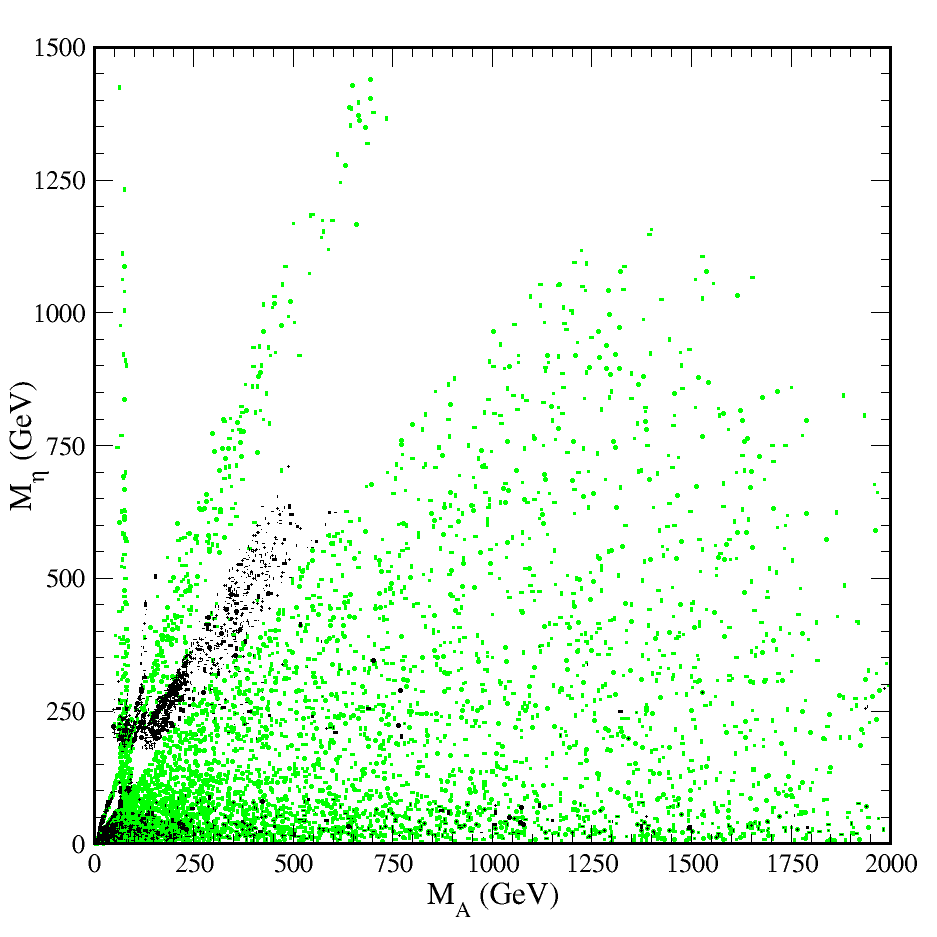}
  \end{minipage}
  \hspace*{-0.2cm}
  \begin{minipage}[t]{0.33\textwidth}
    \centering
    \includegraphics[width=0.985\columnwidth]{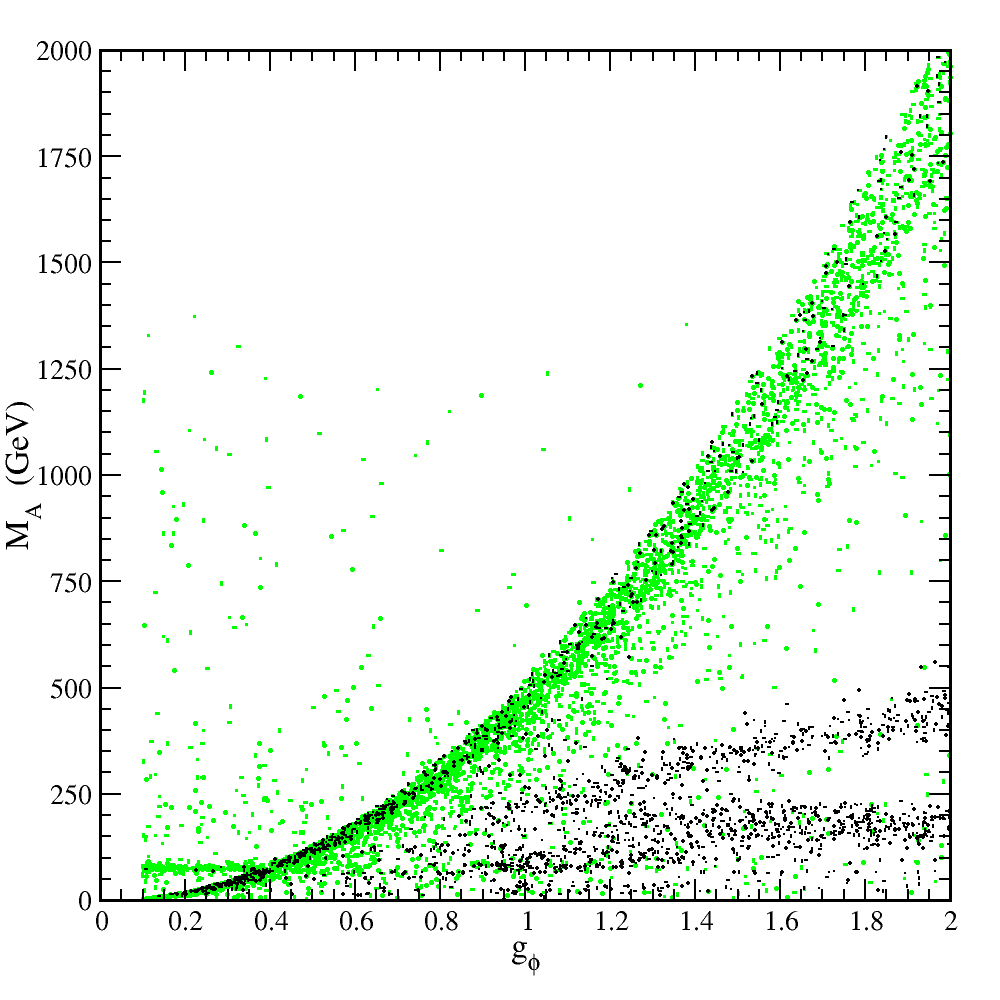}
  \end{minipage}
  \caption{Parameter space leading to $0.091 <\Omega h^2 < 0.129$ for the
  large $\lambda_m$ regime ($10^{-3} < \lambda_m < 1$), from left to right:
  $v_{\phi}$ vs $g_{\phi}$, $M_{\eta}$ vs $M_A$ and $M_A$ vs
  $g_{\phi}$.  $\lambda_{\phi}$ is varied in the range $10^{-5}\-- 1$, the
  Higgs mass between $M_{h}=114.4$ GeV and $M_{h}=160$ GeV. Here too one can recognize the  resonant cases,
 from both the Higgs and the $\eta$ bosons, for $M_A = g_{\phi}v_{\phi}/2
  \sim m_h/2,m_{\eta}/2$ respectively. All dots satisfy the LEP constraints on the $T$ parameter and on $h \rightarrow f\bar{f}$ decay.
They also are in agreement with the direct detection bounds from CDMS \cite{Ahmed:2008eu} and Xenon \cite{Angle:2008we}. The black dots correspond to instances which lead to a spin-independent elastic cross-section at most one order of magnitude below these bounds.}
 \label{fig:llm_ps}
\end{figure}

The corresponding plots for the large Higgs portal regime are given
in Fig~\ref{fig:llm_ps}.
In this case the deviations due to the new $A_i A_j \rightarrow A_k \eta$ processes are more difficult to single out, since
more annihilation channels (involving $\lambda_m$) contribute to the relic density.
But with respect to the case already discussed in
Fig. 4 of Ref.~\cite{Hambye:2008bq}, 
one finds points with lighter $M_{\eta}$ and smaller $g_{\phi}$
for a same value of $v_{\phi}$. The plot $M_A$ vs $g_{\phi}$ indicates
again that the freeze-out has a complicated dependence on the couplings of the
model. Some of the dots still display the quadratic behavior of
Fig.~\ref{fig:slm_ps}, when the dominant annihilation channels are $A_i A_i
\rightarrow \eta \eta$ and $A_i A_j \rightarrow A_k \eta$. 

At tree level the elastic scattering of the vector dark matter on a nucleon $n$ is spin independent, mediated by an $h$ or $\eta$ boson, and the full expression for the cross-section reads:
\begin{equation}
\sigma^{SI} (An \rightarrow An) = \frac{1}{64 \pi^2} f^2 g^4_{\phi} \sin^2 2\beta \ m^2_n \frac{v^2_{\phi}}{v^2}\frac{(M_\eta^2-M^2_h)^2}{M^4_\eta M^4_h}\frac{\mu^2_n}{M_A^2}\,,
\label{sdirdte}
\end{equation}
with $\mu_n=m_n M_A/(m_n+M_A)$ the reduced mass and $m_n$ the nucleon mass. The parameter $f$ designs the Higgs nucleon coupling, $f \equiv \langle n \vert \sum_q m_q \bar{q}q \vert n\rangle$ and is taken to the value of $f=0.3$.

Imposing the relic density constraint, the predictions for the direct detection rate are given in
Fig.~\ref{fig:dirdet}, together with the upper bounds of CDMS~\cite{Ahmed:2008eu}, Xenon10~\cite{Angle:2008we} and the recent final results from CDMS-II~\cite{:2009zw}. 
In the small Higgs portal regime, $\lambda_m \lesssim 10^{-3}$, even though cross sections are suppressed by 2 powers of $\lambda_m$, large direct detection rates can be obtained
for small $\lambda_\phi$ couplings because in this case $m_\eta< m_h$ and the
$A n\rightarrow A n$ cross section scales as $\lambda_m^2/\lambda_\phi^2$, Eq.~(\ref{sdirdte}). 
For a dark matter mass from few GeV all the way up to the multi TeV range, a spin independent cross section of the order of the current experimental sensitivities 
can be obtained
for values of $\lambda_\phi$ of order or below few $10^{-4}$. The values of the various parameters
required in this case can be read off in Fig.~\ref{fig:slm_ps}. 
The value of the cross-section does not
depend on the dark matter mass if this mass is large, as
indicated by the plateaux for different values of $\lambda_{\phi}$. 
For the large Higgs portal regime, $\lambda_m \gtrsim 10^{-3}$, direct
detection rates of order the present experimental sensitivity or exceeding it
are easily produced. For illustration, among the sets of parameters that lead to the right relic density in Fig.~\ref{fig:llm_ps}, we have denoted by black dots the ones which lead to an elastic cross-section on nucleon at most one order of magnitude below the CDMS \cite{Ahmed:2008eu} and Xenon \cite{Angle:2008we} limits.
Here too a dark matter mass in the whole range from 1 GeV up to few TeV can be accommodated. 
Even though other values are possible, the $\eta$ mass tends to be either small, below 100 GeV, or slightly larger than the dark matter mass.
For $M_A$ larger than $\sim 700$ GeV one recovers the linear relation between $v_{\phi}$ and $g_{\phi}$ and the corresponding quadratic behavior of $M_A$ in $g_\phi$, indicating that the pure hidden sector annihilations driven by the $g_\phi$ coupling are dominant, as in the small Higgs portal regime. 

\begin{figure}[t!]
  \begin{minipage}[t]{0.5\textwidth}
    \centering
    \includegraphics[width=0.985\columnwidth]{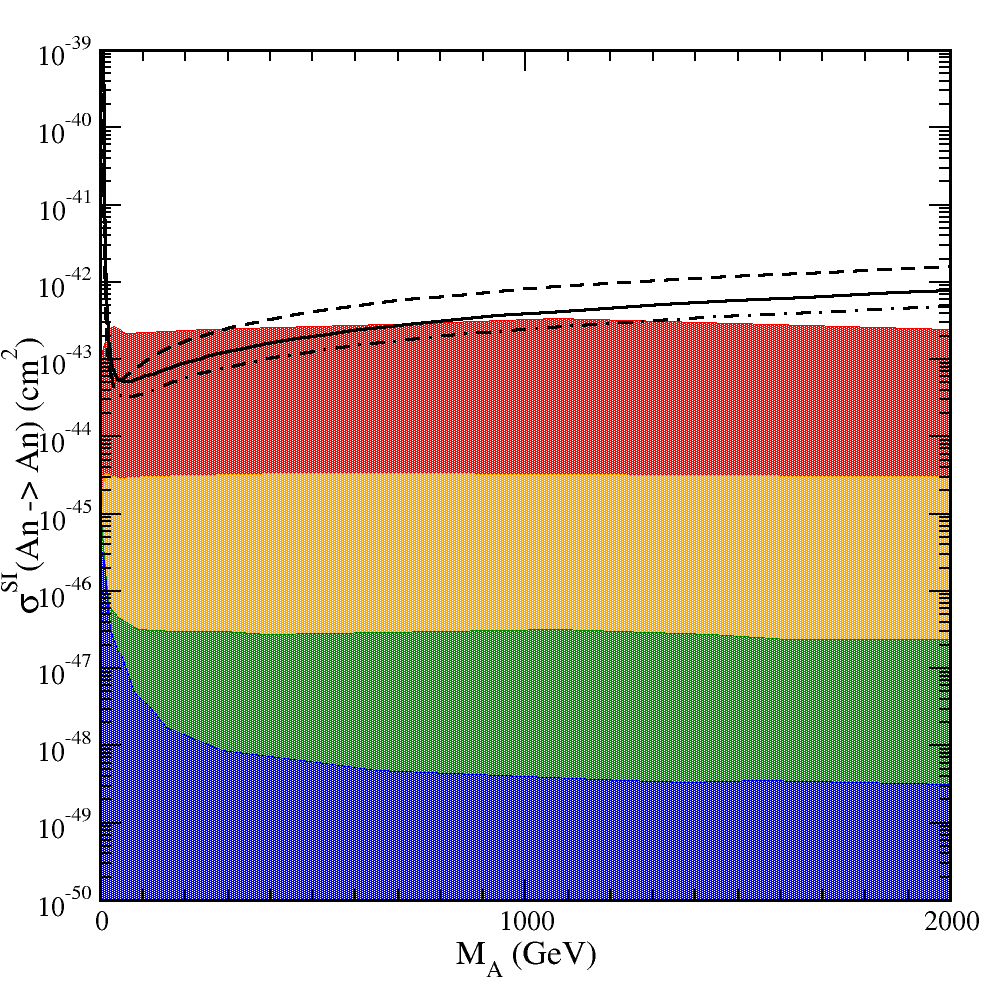}
  \end{minipage}
  \hspace*{-0.2cm}
  \begin{minipage}[t]{0.5\textwidth}
    \centering
    \includegraphics[width=0.985\columnwidth]{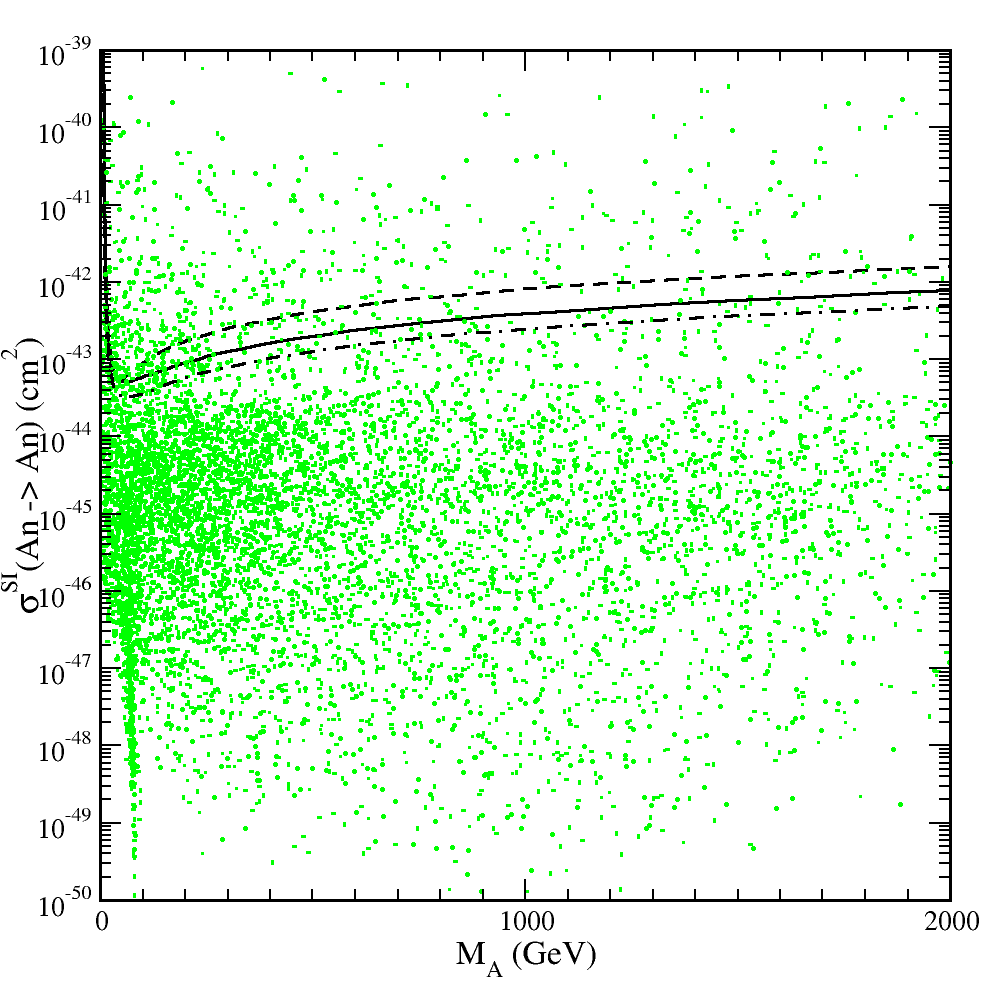}
  \end{minipage}
  \caption{Obtained spin independent cross-section on nucleon $\sigma^{SI} (A n \rightarrow
  A n)$ versus $M_A$, in agreement with the constraint $0.091 <\Omega h^2 < 0.129$. Small
  $\lambda_m$ regime ($10^{-7} < \lambda_m < 10^{-3}$) on the left and large
  Higgs coupling portal, $\lambda_m > 10^{-3}$ on the right. The color caption
  is as in Figs.~\ref{fig:slm_ps} and ~\ref{fig:llm_ps}. The thick (dashed)
  black curve is the CDMS (Xenon10) upper bounds at 90\% C.L.. The dotted-dashed curve is the recent published CDMS-II upper bound, at 90\% C.L.}
  \label{fig:dirdet}
\end{figure}

Note that all the dots shown in the figures 
above satisfy the LEP constraints. The mixing of the $\eta$ boson with the standard
model Higgs affects the electroweak precision observables. The
main constraint on the model parameters comes from the T parameter, since the
$\eta$ is a neutral scalar which mixes with the Higgs boson. We use the same cuts as in
Ref.~\cite{Hambye:2008bq}, that is to require that $T-T_{SM}$ is in the
conservative range $-0.27 \mp 0.05$ from~\cite{lepewwg}. For $M_\eta < 114.4$
GeV the branching ratio $\eta \rightarrow f\bar{f}$ should not exceed the LEP
direct search bounds, leading to an upper value on the mixing angle $\sin^2\beta$, 
see Fig. 10 of Ref~\cite{Barate:2003sz}.
This bound constraints more the large Higgs portal coupling regime which
involve large mixing angles. In addition, we take into account the
combined analysis of the CDF and DO collaborations, that excludes a Higgs in
the mass range $160-170$ GeV at 95\% C.L.~\cite{:2009pt}.

\section{Conclusions}

We have shown in this paper that the vectors of a hidden, spontaneously
broken, non-abelian gauge group constitute a viable dark matter particle
which decays at cosmological times. Their longevity is due to an accidental
custodial symmetry in the renormalizable Lagrangian. However, similarly to the proton,
they are not expected to be absolutely stable due to the existence of non-renormalizable
dimension six operators which induce the decay of the dark matter particle. 
We have identified the four dimension six
operators which break the custodial symmetry and calculated the dominant
decay modes. 
Taking advantage of the fact that the dark matter has spin-1, the gamma lines are produced at tree level from DM two-body decay which beside the $\gamma$ involve the standard model $h$ or hidden sector $\eta$ Higgs bosons.
We have found that in all the cases an intense gamma-ray line is expected,
which could be observed by the Fermi-LAT if the scale of custodial symmetry breaking 
is close to the Grand Unification scale.
We have also calculated the positron fraction, total electron plus
positron flux and the antiproton-to-proton fraction for these channels. 
Even though in these scenarios there is a sizable branching ratio into hadrons, 
the total antiproton-to-proton fraction is consistent with the measurements,
while still producing an observable and possibly huge gamma-ray line. Finally we also
improved the calculation of the relic density of hidden vectors including
the dark matter annihilation processes with one dark matter particle in
the final state. Direct detection rates close to the present limits can easily be produced for any dark matter mass within the GeV-multi TeV range.

\section*{Acknowledgements}
The work of CA and TH is supported by 
the FNRS-FRS, the IISN and the Belgian Science Policy (IAP VI-11).
The work of AI was partially supported by
the DFG cluster of excellence ``Origin and Structure of the Universe.''

\bibliographystyle{JHEP}
\bibliography{bibliography.bib}

\end{document}